\documentclass[pdflatex,sn-mathphys-num]{sn-jnl}

\usepackage[linesnumbered,ruled,vlined]{algorithm2e}
\usepackage{graphicx}
\usepackage{algpseudocode}
\usepackage{multirow}%
\usepackage{amsmath,amssymb,amsfonts}%
\usepackage{amsthm}%
\usepackage{mathrsfs}%
\usepackage[title]{appendix}%
\usepackage{xcolor}%
\usepackage{textcomp}%
\usepackage{manyfoot}%
\usepackage{booktabs}%
\usepackage{listings}%
\usepackage{float}
\usepackage{booktabs}
\usepackage{multirow}
\usepackage{subcaption}
\usepackage{amssymb}
\usepackage{pifont}
\usepackage{caption}
\usepackage{etoolbox}
\usepackage{rotating}
\usepackage{array}

\theoremstyle{thmstyleone}%
\theoremstyle{thmstyletwo}%
\newtheorem{example}{Example}%

\theoremstyle{thmstylethree}%

\begin{document}

\title[SMT-Layout: A MaxSMT-based Approach Supporting Real-time Interaction of Real-world GUI Layout]{SMT-Layout: A MaxSMT-based Approach Supporting Real-time Interaction of Real-world GUI Layout}


\author[1,2]{\fnm{Bohan} \sur{Li}}\email{libh@ios.ac.cn}

\author[3]{\fnm{Dawei} \sur{Li}}\email{davey.lidawei@huawei.com}

\author[3]{\fnm{Ming} \sur{Fu}}\email{ming.fu@huawei.com}

\author*[1,2]{\fnm{Shaowei} \sur{Cai}}\email{caisw@ios.ac.cn}

\affil*[1]{\orgdiv{Key Laboratory of System Software (Chinese Academy of Sciences) and State Key Laboratory of Computer Science}, \orgname{Institute of Software, Chinese Academy of Sciences}, \orgaddress{\state{Beijing}, \country{China}}}

\affil[2]{\orgdiv{School of Computer Science and Technology}, \orgname{University of Chinese Academy of Sciences,}, \orgaddress{\state{Beijing}, \country{China}}}

\affil[3]{\orgdiv{Fields Lab}, \orgname{Huawei Technologies}, \orgaddress{\country{China}}}


\abstract{
  Leveraging the flexible expressive ability of (Max)SMT and the powerful solving ability of SMT solvers, we propose a novel layout model named SMT-Layout.
  SMT-Layout is the first constraint-based layout model that can support real-time interaction for real-world GUI layout adapting to various screen sizes with only one specification.
  Previous works neglect the hierarchy information among widgets and thus cannot exploit the reasoning ability of solvers. For the first time, we introduce Boolean variables to encode the hierarchy relationship, boosting the reasoning ability of SMT solvers. The workflow is divided into two stages. At the development end, two novel preprocessing methods are proposed to simplify constraints and extract useful information in advance, easing the solving burden. After deploying constraints to the terminal end, SMT solvers are applied to solve constraints incrementally. Besides mainstream SMT solvers, a local search solver is customized to this scenario. Experiments show that SMT-Layout can support millisecond-level interaction for real-world layouts, even on devices with low computing power and rigorous memory limitations.
}



\maketitle

\section{Introduction}







Compared to conventional graphical user interface (GUI) layout models such as grid and group, constraint-based layout models are more powerful and flexible:
they can align widgets across the hierarchy~\cite{lutteroth2008modular} and are easier to maintain~\cite{lutteroth2006user,zeidler2012comparing}, and they have been widely applied in GUI design.
For example, Apple's AutoLayout~\cite{sadun2013ios} adapts interfaces on different devices.
CSS's Flex Box uses constraints to fit the content and solve alignment problems~\cite{weyl2017flexbox}.

The early typical constraint-based layout models are based on linear constraint systems. The specifications of layout are modeled as linear equalities and inequalities, and solved by linear or quadratic programming solvers ~\cite{badros2001cassowary, bill1992bricklayer, borning1997solving, hosobe2000scalable, lutteroth2008domain}.
However, constraint-based layout models share the same drawback as conventional layout models: when the layouts need to adapt to screens with drastically different sizes, multiple specifications need to be defined and synchronized, which can be time-consuming and can cause maintenance difficulties.


In 2019, a constraint-based layout model called ORCLayout was proposed to specify layout based on OR constraints~\cite{jiang2019orc}, allowing the GUI layout to adapt to multiple screen sizes with only one specification. 
An OR constraint is a disjunction of constraints where only one needs to be true, and soft constraints with weights are introduced to express the preference of different terms in an OR constraint. 
In essence, ORCLayout models the layout problem as an optimized version of {\it Satisfiability Modolo Theories (SMT)}, also known as MaxSMT.
(Max)SMT is an important fragment of the formal method.

However, state-of-the-art (Max)SMT solvers such as Z3 perform poorly on ORCLayout constraints, mainly because ORCLayout cannot model the hierarchy relationships among widgets with Boolean constraints, and thus the powerful reasoning ability of SMT solvers cannot be utilized to reduce the exponent combination of OR constraints.
A solver called ORCSolver is proposed to solve the MaxSMT problem, based on a branch-and-bound (B\&B) approach with modular heuristics dedicated to specific layout patterns~\cite{jiang2020orcsolver}, which is not a general method and does not support incremental SMT solving.
Overall, ORCLayout does not leverage the main power of SMT solvers including reasoning and incremental solving, and thus it cannot achieve real-time interaction.

Moreover, ORCLayout cannot handle real-world layout design since it does not fully utilize the expression ability of (Max)SMT.
Specifically, it does not provide abundant layout patterns supporting practical design.




As summarized in Table~\ref{layout_cmp}, the conventional layout model and linear constraint-based layout model can support real-time interaction for real-world layouts, and they have been widely applied in practical GUI design.
However, they need to maintain multiple specifications for various screen sizes, which can be time-consuming and may cause consistency issues.
In contrast, ORCLayout can adapt to multiple screen sizes with one unique specification, but it is not suitable for practical GUI design because it cannot handle real-world layout and cannot support real-time response.
As far as we know, no layout model can support real-time interaction for real-world GUI layouts adapting various screen sizes with only one specification.

In this paper, by fully utilizing the flexible expression ability of (Max)SMT and the powerful solving ability of SMT solvers, 
we aim to propose a practical constraint-based layout model named SMT-Layout to satisfy all dimensions in Table~\ref{layout_cmp}.

\begin{table}[]
\caption{Comparison of Different Layout Models}
\label{layout_cmp}
\begin{tabular}{@{}llll@{}}
\toprule
                        & Real-world Layout  & Real-time Interaction & Unique specification \\ \midrule
Conventional Layout      & \checkmark                                  & \checkmark     & \ding{55}           \\
Linear Constraint-based Layout & \checkmark                                  & \checkmark        & \ding{55}        \\
ORCLayout          & \ding{55}                                 & \ding{55}          & \checkmark        \\
SMT-Layout          & \checkmark                                & \checkmark        & \checkmark         \\ \bottomrule
\end{tabular}
\end{table}



\subsection{Contributions}

We propose a practical modeling module based on the flexible expression ability of MaxSMT:
abundant preset layout containers are provided for designers to flexibly specify real-world layouts and convert them to MaxSMT constraints.
A special layout container named ``Placeholder'' is proposed to accommodate alternative layouts for varying screen sizes with only one specification.






We consider the solving efficiency to be the main bottleneck since solving the layout specifications in the form of MaxSMT is NP-hard.
As shown below, the main effort is devoted to achieving real-time interaction as the screen size varies, leveraging the reasoning and incremental solving ability of SMT solvers.
Note that our approaches are general and do not rely on specific layout patterns, so they can adapt to new layout patterns without special customization.

This work introduces Boolean variables to illustrate the visibility property of widgets, and for the first time, directly specifies the hierarchy relationship among widgets by the Boolean constraints on their visibility property.
Based on these Boolean constraints, the reasoning ability of the SMT solver can be fully utilized to infer invisible widgets and omit their corresponding constraints, which can drastically simplify the original formula, thus boosting the solving efficiency.
Soft constraints are also introduced as Boolean variables with corresponding weights to express the preference.

We propose to ease the solving burden by simplifying the formulas and extracting useful information in advance.
Specifically, a two-stage architecture is proposed to divide the workflow into the {\it development end} and the {\it terminal end}.
The layout is specified and converted to formulas at the development end, and the converted formulas are then deployed to the terminal end for solving and display.

At the development end, two novel preprocessing methods are proposed to enhance the solving efficiency:
First, the {\it Interval-based soft constraints hardening} method converts the MaxSMT formula to SMT formulas by predetermining the assignments of soft constraints corresponding to the interval of the size property.
With these predetermined Boolean variables, we can apply the reasoning ability of SMT solvers to infer the visibility property of widgets, based on the Boolean constraints w.r.t. the hierarchy relationship.
The corresponding constraints regarding those invisible widgets can be omitted, significantly reducing the search space.
Second, the {\it Independent widget extraction} method utilizes the modular nature of layout designing to extract independent widgets, and the original formula is divided into multiple independent formulas with moderate scales, which are easier to solve.

After the preprocessing, the preview module evaluates the layout in advance and prompts the designers about potential issues.

At the terminal end, a two-level solving strategy is proposed to solve these constraints, leveraging the reasoning and incremental solving power of SMT solvers.
Besides the mainstream complete SMT solver such as Z3~\cite{moura2008z3}, a lightweight local search SMT solver called LocalSMT is customized as the backend solver.
Local search is appropriate for the incremental scenario, since most constraints remain satisfied when resizing the screen, and local search can be applied to efficiently find a new solution near the current one.
Moreover, LocalSMT has a compact data structure and can save memory usage, which is suitable for the terminal end with rigorous memory limitations.

Overall, we apply the main power of (Max)SMT to design and implement a novel constraint-based layout model called SMT-Layout, based on the following contributions:

\begin{itemize}
    \item Based on the expression ability of MaxSMT, a novel modeling module is proposed to specify the GUI layout.
    Abundant layout containers are provided to support real-world layouts adapting to various screen sizes with only one specification (Sect.~\ref{preset container}).
    \item Boolean variables are introduced to illustrate the visibility property of widgets, based on which the hierarchical relationships are directly specified with Boolean constraints, boosting the reasoning ability of SMT solvers (Sect.~\ref{boolean hierarchy relation}).
    \item A two-stage workflow architecture is proposed to ease the solving burden at the terminal end (Sect.~\ref{archi}).
    Specifically, at the development end, two novel preprocessing methods, namely {\it Interval-based soft constraints hardening} and {\it Independent widget extraction}, are proposed to simplify the formulas and extract useful information in advance (Sect.~\ref{preprocess module}).
    \item A two-level solving strategy is proposed to solve the constraints based on the reasoning and incremental solving ability of SMT solvers (Sect.~\ref{two-level}).
    Besides mainstream complete SMT solvers, a local search SMT solver is customized for the incremental scenario as the backend solver, enhancing the solving efficiency and saving memory usage (Sect.~\ref{sls for layout}).
\end{itemize}

Based on SMT-Layout, we have specified 12 layouts derived from real-world GUIs, and each of them can adapt to various screen sizes with only one specification. 
The experimental results show that SMT-Layout can achieve millisecond-level real-time responsiveness in handling these layout specifications, even on devices with constrained computational resources, such as mobile phones.
Moreover, the memory overhead is minimal with local search solvers serving as the backend solver (Sect.~\ref{evaluation}).


\section{Preliminary}
\label{pre}
\subsection{SMT, MaxSMT, OMT}
As an important fragment of the formal method, Satisfiability Modulo Theories (SMT) decides the satisfiability of a first-order logic formula concerning certain background theories.
Specifically, in the context of constraint-based layout, the background theory is the linear arithmetic theory, consisting of arithmetic formulae in the form of linear equalities or inequalities over arithmetic variables ($\sum_{i=0}^n{a_ix_i\leq k}$ or $\sum_{i=0}^n{a_ix_i = k}$, where $a_i$ are coefficients and $x_i$ are arithmetic variables).
An atomic formula can be a propositional variable or an arithmetic formula.
A $literal$ is an atomic formula, or the negation of an atomic formula.
A $clause$ is the disjunction of a set of literals, and a formula in {\it conjunctive normal form (CNF)} is the conjunction of a set of clauses.
A $clause$ consisting of only one literal is defined as {\it unit clauses}.

The MaxSAT modulo theories (MaxSMT) problem is an optimization version of SMT, where clauses are divided into {\it hard} clauses and {\it soft} clauses with weight.
MaxSMT aims to maximize the total weight of satisfied soft clauses while satisfying all hard clauses.
Optimized Modulo theories (OMT) is an extension of SMT with an objective.
OMT aims to find models that optimize given objectives while satisfying all constraints.

The core SMT solving builds upon the strengths of SAT~\cite{kroening2016decision,ganzinger2004dpll}, which contains the following Boolean automated reasoning techniques~\cite{biere2009handbook}:
{\it Unit Propagation}, {\it Pure Literal Elimination}, {\it Variable Elimination} and {\it Subsumption Elimination} etc.
These techniques can efficiently handle the Boolean constraints and drastically simplify the SMT formula.
However, in previous works, Boolean variables were not introduced to the constraint-based layout models, so the Boolean automated reasoning ability of SMT solvers was not fully utilized.

\subsection{ Local Search SMT solvers}
Local search is an incomplete method that has recently been successfully applied to solve the SMT problem~\cite{frohlich2015stochastic,niemetz2016precise,niemetz2017propagation}.
LocalSMT~\cite{cai2023local,cai2022local,li2023local} is the state-of-the-art local search solver dedicated to SMT with the linear arithmetic theory.
It iteratively modifies the current solutions by applying the ``{\it critical move}'' operator, which is appropriate for incremental scenarios.
Specifically, the {\it critical move} operator assigns an arithmetic variable $x$ to the threshold value making constraint $\ell$ true, where $\ell$ is a falsified constraint containing $x$.
Moreover, LocalSMT is a lightweight solver with a compact data structure, so it is suitable for scenarios with rigorous memory restrictions.

\section{Overview of Two-stage Workflow Architecture}
\label{archi}



\begin{figure}[h]
  \centering
  \includegraphics[width=\linewidth]{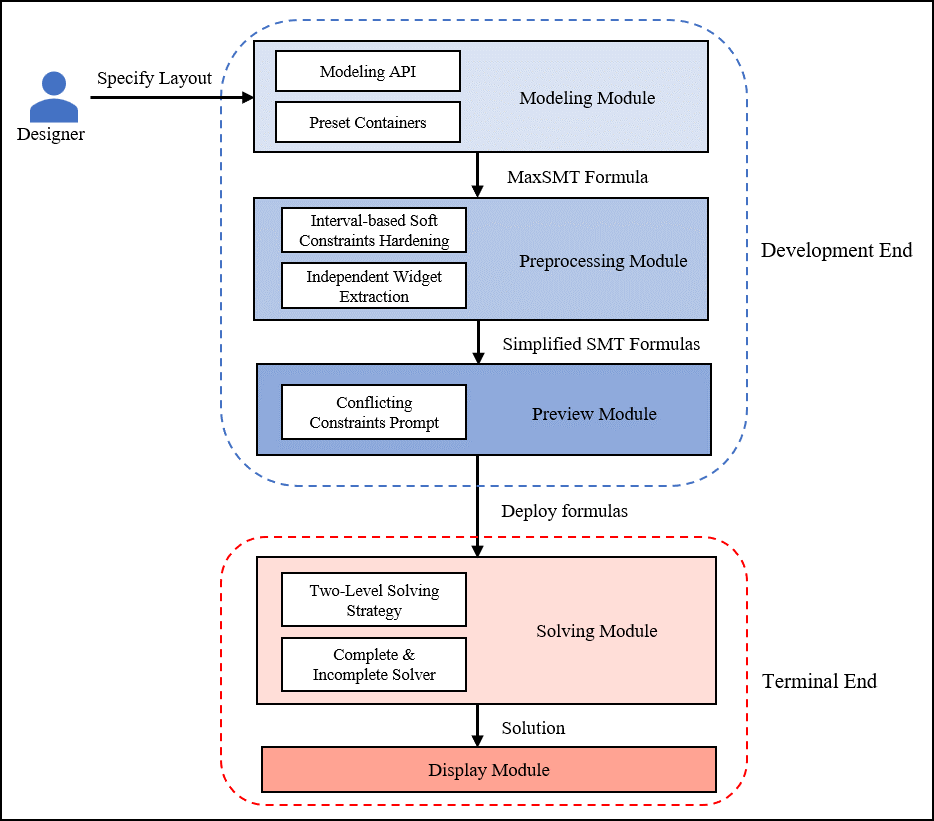}
  \caption{The two-stage workflow architecture of SMT-Layout.}
  \label{framework}
\end{figure}

As shown in Fig.~\ref{framework}, to support real-time interaction of layouts even on terminal devices with limited computing resources, we propose to divide the workflow into two stages, namely the {\it development end} and the {\it terminal end}.
Compared to the terminal end, the development end has sufficient computational resources, which can be utilized to ease the solving burden at the terminal end.

{\bf Development End: }
First, in the modeling module, the modeling Application Programming Interface (API) and preset layout containers are provided for designers to specify the layouts and convert the specifications to MaxSMT formulas.
By leveraging the abundant computational resources at the development end, two novel preprocessing methods, namely {\it Interval-based soft constraints hardening} and {\it Independent widget extraction}, are proposed to simplify the formula and extract useful information in advance to enhance solving efficiency.
Subsequently, the preview module evaluates whether the layout meets expectations and notifies designers of potential issues.

{\bf Terminal End: }
Afterward, the simplified formulas and extracted information are deployed to the terminal end.
In the solving module, a two-level solving strategy is proposed to solve the constraints based on the current screen size, leveraging the reasoning and incremental solving power of SMT solvers.
In addition to complete SMT solvers such as Z3~\cite{moura2008z3}, a lightweight local search SMT solver is customized as the backend solver for this incremental scenario to improve the solving efficiency and reduce memory usage.
Finally, the solution is transferred into the layout for display.

Note that, since the layouts of most practical web pages and mobile Apps support vertical scrolling, the impact of screen height on layout is minimal, and screen width becomes the primary factor influencing layout.
Therefore, SMT-Layout focuses on the real-time interaction of real-world layouts in response to variations in screen width.

\section{Modeling Module}
\label{model}
This section describes the modeling module to formalize the layout based on MaxSMT.
First, we define the basic properties of widgets, introducing the Boolean variable to illustrate the $visibility$ of widgets.
Then, abundant preset layout containers are provided to specify real-world layouts, and a novel container called {\it Placeholder} is proposed to support layout alternatives adapting to various screen sizes with one specification.

Besides numerical constraints illustrating the positional relationships, for the first time, Boolean constraints are applied to directly specify the hierarchical relationships among widgets as hard constraints.
Based on these Boolean constraints, the reasoning capability of SMT solvers can be utilized to infer the visibility of widgets in advance, and the constraints regarding invisible widgets can be omitted to reduce the search space.
Finally, soft constraints are introduced as Boolean variables to express the preference for alternative layouts.

\subsection{Basic properties of widgets}
\begin{figure}[h]
  \centering
  \includegraphics[width=0.35\linewidth]{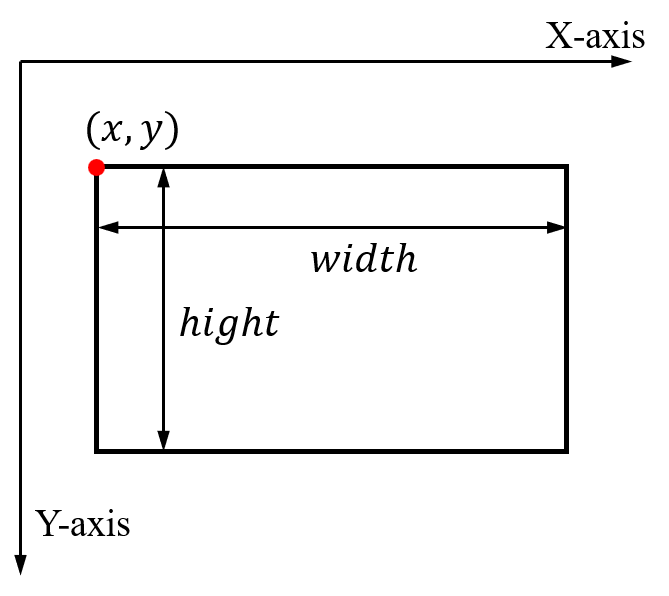}
  \caption{The numerical properties of widgets.}
  \label{property}
\end{figure}

As shown in Fig.~\ref{property}, each widget can be viewed as a rectangle with 4 numerical properties: the size properties namely $width$ and $height$; the position properties illustrating the coordinates of the upper left corner, namely $x$ and $y$.

Moreover, we propose associating each widget with a Boolean property to determine whether the widget is visible, denoted as $visibility$.
This Boolean property will be applied to illustrate the hierarchy relationships among widgets in the following context.
If a widget is invisible, it will not be displayed and its corresponding constraints will not take effect.

Given a widget $w$, these five properties are denoted as $[w]_w$, $[w]_h$, $[w]_x$, $[w]_y$ and $[w]_v$ respectively.
After solving the constraints, the solution consists of these properties, and visible widgets will be displayed according to their size and position properties.

\subsection{Preset Containers}
\label{preset container}
Abundant typical layout patterns, such as {\it Row}, {\it Column}, {\it Card}, {\it Table}, {\it Flow}, {\it Waterfall}, and {\it Flex} are provided as preset containers to specify real-world layouts.
Containers provide a structural framework for organizing the widgets, and designers can apply containers to specify the layout hierarchically.
Various position constraints among widgets are supported through numerical constraints, including alignment, centering, adjacency, etc.
Default position constraints are automatically generated and maintained for containers.
Moreover, designers can flexibly employ the API to add individual constraints for fine-tuning the widgets.
Detailed implementations and formalization of these preset containers are provided in the Appendix~\ref{preset_container}.

In addition to typical layout containers, a special container called {\bf Placeholder} is proposed to accommodate alternative layouts for varying screen sizes.
The {\it Placeholder} container contains multiple widgets that cannot simultaneously appear, and these widgets will switch as the screen size varies.
\begin{figure}[h]
  \centering
  \includegraphics[width=0.65\linewidth]{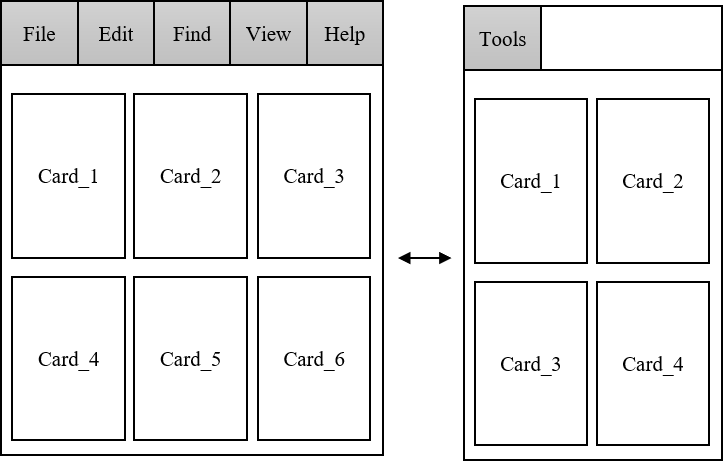}
  \caption{Alternative layouts with varying screen size: The range of screen width is $[1000,2000]$. The wide version on the left is presented if the screen width is in the interval $[1500,2000]$.
    Otherwise, the compact version on the right is presented in the interval $[1000,1500)$.}
  \label{fig_ex_card}
\end{figure}

\begin{figure}[h]
  \centering
  \includegraphics[width=0.65\linewidth]{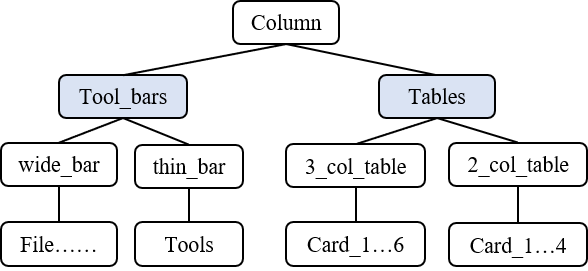}
  \caption{The corresponding hierarchical relationship. Blue item item denotes the {\it Placeholder} containers.}
  \label{fig_ex_relation}
\end{figure}

\begin{example}
\label{exp_card}
As shown in Fig.~\ref{fig_ex_card}, the layout is modified as the screen width varies, and the hierarchical relationship is presented in Fig. \ref{fig_ex_relation}.
Two {\it Placeholder} containers, namely $Tool\_bars$ and $Tables$ are arranged in a {\it Column} container.
The $Tool\_bar$ contains 2 alternative {\it Row} containers, namely $thin\_bar$ and $wide\_bar$, while the $Tables$ contains 2 alternative {\it Table} containers, namely $3\_col\_table$ and $2\_col\_table$, containing multiple {\it Card} widgets, arranged in 3 columns and 2 columns respectively.
\end{example}

\subsection{Hierarchy Relationships Specified by Boolean Constraints}
\label{boolean hierarchy relation}
In addition to the numerical constraints illustrating positional relationships, we propose employing Boolean constraints to specify the hierarchy relationship among widgets based on their $visibility$ property, which can boost the reasoning capability of SMT solvers to infer invisible widgets in advance, thus omitting their corresponding constraints and reducing search space.

{\bf Hierarchy relationship:} Since containers are organized hierarchically, for each widget $w$, the set of containers to which it belongs is defined as $parents$ of $w$, denoted as $[w]_p$.
Conversely, the set of widgets contained within $w$ is defined as $kids$ of $w$, denoted as $[w]_k$.
Moreover, we denote the set of widgets in the sub-layout it contains as $[w]_{sub}$, which can be inferred by recursively traversing its $kids$~\footnote{$[w]_{sub}$ is the superset of $[w]_k$. For example, in Fig.~\ref{fig_ex_relation}, $[Tool\_bars]_k$ contains $wide\_bar$ and $thin\_bar$, while $[Tool\_bars]_{sub}$ contains all widgets in the subtree of $Tool\_bars$.}.
Note that, to facilitate the modular reuse of existing specifications, a widget can be contained in multiple containers, indicating that the $parents$ of a widget may contain multiple widgets.
For example, the $parent$ of the widget $Card\_1$ includes $3\_col\_table$ and $2\_col\_table$.

{\bf Boolean constraints:} As shown below, the Boolean constraints define the hierarchical relationships among widgets based on their $visibility$ properties, which are automatically generated and maintained.

If a widget $w$ is visible, there is only one visible widget among the $parents$ of $w$, since it cannot simultaneously appear in multiple containers. For example, if $Card\_1$ is visible, then either $3\_col\_table$ or $2\_col\_table$ is visible, since $Card\_1$ cannot simultaneously appear in both Table containers:
\begin{equation}
    [w]_v\rightarrow(\bigvee_{p_i\in [w]_p}[p_i]_v) \quad \texttt{(at least one widget in $[w]_p$ is visible)}
\end{equation}

\begin{equation}
    \forall p_i,p_j\in[w]_p,p_i\not = p_j,[w]_p\rightarrow(\neg[p_i]_v\vee\neg[p_j]_v) \quad \texttt{(at most one widget in $[w]_p$ is visible)}
\end{equation}

If a typical container $w$ is visible, all widgets among its $kids$ are visible. For example, if $3\_col\_table$ is visible, then all of its kid widgets, namely $Card\_1$ to $Card\_6$ should be visible:
\begin{equation}
    [w]_v\rightarrow(\bigwedge_{k_i\in[w]_k}[ k_i]_v)
\end{equation}

As a special case, the $visibility$ properties of widgets among the $kids$ of the non-wrap version of the {\it Flow} layout are determined by their position, details can refer to the implementation of the {\it Flow} layout in Appendix~\ref{flow_layout}.

Given a {\it Placeholder} container $w$, if it is visible, there is only one visible widget among the $kids$ of $w$, since these alternative layouts will switch between each other. For example, if the Placeholder $Table$ is visible, then either $3\_col\_table$ or $2\_col\_table$ should be visible:
\begin{equation}
[w]_v\rightarrow(\bigvee_{k_i\in [w]_k}[ k_i]_v)\quad \texttt{(at least one widget in $[w]_k$ is visible)}
\end{equation}

\begin{equation}
    \forall k_i,k_j\in[w]_k,k_i\not = k_j,[w]_p\rightarrow(\neg[ k_i]_v\vee\neg[ k_j]_v)\quad \texttt{(at most one widget in $[w]_k$ is visible)}
\end{equation}

Moreover, given a container $w$, the associated positional constraints regarding itself and its kid widgets, denoted as $position\_c(w)$, will take effect when the container is visible.

\begin{equation}
    [w]_v\rightarrow position\_c(w)
\end{equation}

Thus, the reasoning ability of SMT solvers can be applied to infer the $visibility$ property of widgets, and those positional constraints associated with invisible widgets can be omitted, drastically reducing the search space.
As the example shown in Fig.~\ref{fig_ex_card}, the alignment constraints among the {\it Card} widgets in the $3\_col\_table$ container will take effect when $3\_col\_table$ is visible.
In the compact version where $3\_col\_table$ is invisible, the corresponding positional constraints can be omitted.

\subsection{Soft Constraints}

Besides hard constraints including numerical constraints illustrating the positional relationship and Boolean constraints illustrating the hierarchy relationship, soft constraints are introduced to express preferences for alternative layouts. 
The soft constraints are in the form of Boolean variables with associated weights, and the goal is to maximize the total weight of satisfied soft constraints while satisfying all hard constraints.
These soft constraints can be divided into the following 2 types:

First, designers can flexibly define preferences for positional relationships among widgets, by introducing auxiliary Boolean variables as soft constraints.
Specifically, the positional relationship is formalized as the conjunction of constraints.
For example, the positional relationship that widgets $a$ and $b$ are center-aligned is formalized as $(2*[a]_x+[a]_w =2*[b]_x+[b]_w)\wedge(2*[a]_y+[a]_h =2*[b]_y+[b]_h) $.
For each conjunction $F_{conj}$, the preference can be specified by introducing a fresh auxiliary Boolean variable $soft\_new$  and adding the following hard constraints:
\begin{equation}
    soft\_new\rightarrow F_{conj}\quad \texttt{(if $soft\_new$ is true, $F_{conj}$ should be satisfied)}
\end{equation}

These auxiliary variables are set as soft constraints, indicating that they are preferred to be true, and consequently, the corresponding $F_{conj}$ are also preferred to be satisfied.

Second, given a $Placeholder$ container, the $visibility$ properties of its alternative widgets are set as soft constraints with different weights, expressing different preferences for these alternative widgets.
The set of these soft constraints can be formalized as:
\begin{equation}
    \{[w]_v|w\in [p]_k, p\;is\; a\;Placeholder \}
\end{equation}

As shown in the example in Fig. \ref{fig_ex_card}, the soft constraints are the $visibility$ property of the alternative widgets.
In the wide version, the truth assignment of soft constraints is $\{ [wide\_bar]_v$, $\neg[thin\_bar]_v$,$[3\_col\_table]_v$,$\neg[2\_col\_table]_v\}$. 
In the compact version, the truth assignment of soft constraints is  $\{ \neg[wide\_bar]_v$, $[thin\_bar]_v$,$\neg[3\_col\_table]_v, [2\_col\_table]_v\}$.

Given a MaxSMT formula $F_{max}$, the corresponding hard and soft constraints are denoted as $F_{max}.hard$ and $F_{max}.soft$ respectively.


\section{Preprocessing Module}
\label{preprocess module}
After the modeling module, the specification of layouts is converted to MaxSMT formulas.
However, solving MaxSMT formulas can be time-consuming in practice and cannot support real-time interaction at the terminal end.
We leverage the abundant computational resources at the development end to enhance the solving efficiency.
Hence, two novel preprocessing methods are proposed to simplify the original formula and extract useful information in advance to guide the search.

\subsection{Interval-based Soft constraints hardening}

We observe that the layout structure remains consistent within a specific interval of screen width, indicating that the truth assignments of soft constraints remain constant within such intervals.
Furthermore, once the truth assignments of soft constraints are determined, the MaxSMT formula actually becomes the SMT formula.
This conversion can improve solving efficiency, as SMT solving has made significant progress.
Finally, once the $visibility$ properties of alternative widgets are determined, SMT solvers can automatically reason the $visibility$ of other widgets, depending on the Boolean constraints that illustrate the hierarchical information.
The formula can be simplified by excluding the constraints associated with the invisible containers.

Based on the above observation, a preprocessing method called {\it Interval-based soft constraints hardening} is proposed to fully utilize the efficient solving ability and automated reasoning ability of SMT solvers. 
Given a size property $p$ and its related soft constraints, the method gradually traverses the range of $p$, determining the truth assignments of soft constraints corresponding to the interval of $p$.
Their correspondence relationships are detected as constraints, which can be used to convert MaxSMT formulas into SMT formulas.
Specifically, the method is comprehensively described in Alg. \ref{soft_harden_alg}, as outlined below:

\begin{algorithm}[!t]
\caption{ soft\_constraints\_hardening}
\label{soft_harden_alg}
\SetKwInOut{Input}{INPUT}
\SetKwInOut{Output}{OUTPUT}
\Input{A MaxSMT formula $F_{max}$ and the size property $p$}
\Output{$C\_relation$, the constraint formalizing the relationship between the intervals of $p$ and the corresponding truth assignments of soft constraints}
find the maximum and minimum value of $p$, $max\_val$ and $min\_val$, by applying OMT solver to $F_{max}.hard$ \; 
$C\_relation:=(min\_val\le p\le max\_val);${\tcc {determine the range of $p$}}
$upper\_bound:= max\_val$\;
\While{$upper\_bound>min\_val$}{
find the solution $\alpha$ by applying MaxSMT solver to solve $F_{max}\wedge(p=upper\_bound)$\;
$\alpha_{soft}:=$ the truth assignment of soft constraints related to $p$ at the current size in $\alpha$\;
find  $lower\_bound$, the minimum value of $p$ regarding $\alpha_{soft}$, by applying OMT solver to solve $F_{max}.hard \wedge (\bigwedge_{\ell\in \alpha_{soft}}\ell)$\;
$C\_relation:= C\_relation\wedge((\bigwedge_{\ell\in \alpha_{soft}}\ell)\leftrightarrow (lower\_bound\le p\le upper\_bound))$\;
{\tcc {formalize the correspondence relationship with constraint}}
$upper\_bound := lower\_bound-1$
}
{\bf return} $C\_relation$\;
\end{algorithm}

First, we determine the maximum and minimum values of the size property $p$, namely $max\_val$ and $min\_val$, by adopting the OMT solver to solve the hard constraints of the MaxSMT formula (Line 1).
The size property is limited in the range of $[min\_val, max\_val]$ (Line 2).
Then, we traverse the range of $p$ and divide it into intervals based on the assignment of soft constraints related to $p$, denoted as $\alpha_{soft}$ (Lines 3--9).
The upper bound of the current interval, denoted as $upper\_bound$, is initialized as $max\_val$ (Line 3).
The MaxSMT solver is employed to solve the MaxSMT formula attached with the constraint $(p = upper\_bound)$ to obtain the solution $\alpha$ (Line 5), which is used to extract $\alpha_{soft}$ at the current size (Line 6).
Then we determine the minimum possible value of $p$ regarding $\alpha_{soft}$, denoted as $lower\_bound$.
It is accomplished by applying the OMT solver to solve the hard constraints attached with $(\bigwedge_{\ell\in \alpha_{soft}}\ell)$, indicating that all soft constraints are fixed to the truth value specified in $\alpha_{soft}$ (Line 7).
$[lower\_bounnd, upper\_bound]$ represents the feasible interval of $p$ corresponding to $\alpha_{soft}$.
Their relationship is formalized as $(\bigwedge_{\ell\in \alpha_{soft}}\ell)\leftrightarrow (lower\_bound\le w\le upper\_bound)$, indicating that the truth assignment of soft constraints is set as $\alpha_{soft}$ if and only if $p$ falls into this interval (Line 8).
The value of $upper\_bound$ decreases to continue searching for the next interval (Line 9) until it reaches $min\_val$ (Line 3).
Finally, the algorithm outputs the constraint $C\_relation$, representing the relationship between the truth assignments of soft constraints and their corresponding intervals of $p$ (Line 10).

Applying the preprocessing method to the screen width $[screen]_w$, the range of screen width is divided into multiple intervals, and within each interval, the corresponding truth assignments of soft constraints are determined.
The relationships are formalized as the constraint $C\_relation$.
By substituting the soft constraints with the resulting constraint, the MaxSMT formula is transformed into an SMT formula, which can be efficiently solved by SMT solvers.
Furthermore, since soft constraints consist of the $visibility$ properties of alternative widgets, once they are determined by identifying the interval to which the current screen width belongs, the SMT solvers can infer the $visibility$ of other widgets based on Boolean constraints derived from the hierarchical information, leveraging the reasoning power.
Consequently, the constraints of those invisible containers can be omitted, thus reducing the search space.

\begin{example}
    As shown in Fig. \ref{fig_ex_card}, after applying the {\it Interval-based soft constraints hardening} preprocess on the screen width property, the range of screen width is determined as $[1000,2000]$, which is formalized as $(1000\le [screen]_w\le2000)$, and can be divided into 2 intervals: $[1500,2000]$ and $[1000,1500)$.
    The relationships between these intervals and the corresponding truth assignments of soft constraints are formalized as follows: $([wide\_bar]_v\wedge \neg [thin\_bar]_v \wedge[3\_col\_table]_v\wedge\neg [2\_col\_table]_v)\leftrightarrow(1500\le [screen]_w\le 2000)$, and $(\neg [wide\_bar]_v\wedge [thin\_bar]_v\wedge\neg [3\_col\_table]_v\wedge[2\_col\_table]_v)\leftrightarrow(1000\le [screen]_w<1500)$ respectively.
    These constraints are integrated into the original hard constraints, converting the MaxSMT formula to an SMT formula.
    
\end{example}


{\bf Insight: }The {\it Interval-based soft constraints hardening} preprocessing method can be regarded as detecting the intrinsic constraints of the MaxSMT formula on the size property $p$.
Specifically, the hard constraints determine the range of $p$,
while the preference for soft constraints is reflected in the relationship between the interval of $p$ and the corresponding truth assignment of soft constraints.

\subsection{Independent Widgets Extraction}

In the practical layout development process, the widgets are designed and assembled in a modular manner.
Thus, the sub-layouts of certain widgets do not affect other widgets outside the sub-layouts, and they are considered as ``independent'' widgets.
By extracting the constraints related to ``independent'' widgets, the original formula is divided into a simplified overall formula and multiple formulas specifying the sub-layouts of ``independent'' widgets.
Compared to the original formulas, these formulas are smaller and can be solved more efficiently.

The criteria for the ``independent'' widget is formally defined as follows:
given a widget $w$, the set of clauses involving the properties of any widget in $[w]_{sub}$ is defined as the $sub\_constraints$ of $w$, denoted as $F_{sub}(w)$,
and $w$ is {\bf independent} if each clause $c\in F_{sub}(w)$ does not concern the properties of any widget $w'\not \in \{w\cup[w]_{sub}\}$.
In other words, the $sub\_constraints$ of an independent widget does not affect any widget outside the sub-layout, and it only affects the sub-layout itself.

\begin{figure}[]
    \centering
    \begin{subfigure}[b]{0.45\linewidth}
      \centering
      \includegraphics[width=4.5cm]{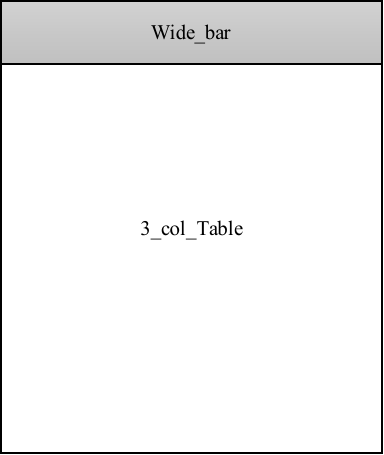}
      \caption{Outer layer: abstraction}
      \label{outer}
    \end{subfigure}
    \begin{subfigure}[b]{0.45\linewidth}
    \centering
      \includegraphics[width=4.5cm]{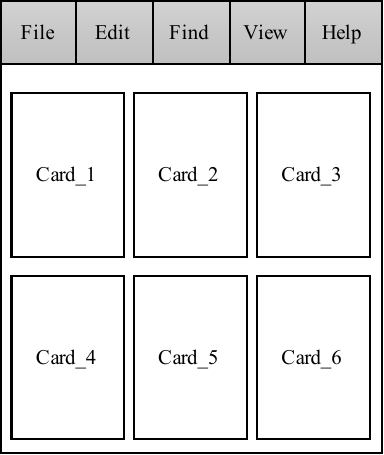}
      \caption{Inner layer: refinement}
      \label{refine}
    \end{subfigure}
    \caption{Independent widgets Extraction: $Wide\_bar$ and $3\_col\_Table$ are independent widgets, since their $sub\_constraints$ do not affect widgets outside their sub-layouts, and can be extracted. In the solving module, (a) the overall layout in the outer layer is determined at the abstraction level, and (b) the sub-layouts in the inner layer are refined at the refinement level.}
    \label{extract}
\end{figure}

\begin{algorithm}[!t]
\caption{ Independent widgets extraction}
\label{inner_outer}
\SetKwInOut{Input}{INPUT}
\SetKwInOut{Output}{OUTPUT}
\Input{A MaxSMT formula $F_{max}$}
\Output{the overall constraints in the outer layer $Outer\_constraint$, and the set of constraints for independent widgets in the inner layer $Inner\_constraints$}
$Inner\_constraints:=\phi$\;
Extract the $sub\_cosntraints$ of independent widgets from $F_{max}$\;
\For{$w$ in independent widgets}{
$C_{width}:=$ soft\_constraints\_hardending($F_{sub}(w)$,$[w]_w$)\;
$C_{height}:=$ soft\_constraints\_hardending($F_{sub}(w)$,$[w]_h$)\;
{\tcc{detect the influence of $F_{sub}(w)$ on $[w]_w$ and $[w]_h$}}
$C\_inner:=F_{sub}(w).hard\wedge C_{width}\wedge C_{height};$  {\tcc{convert $F_{sub}(w)$ to inner-layer SMT formula}}
$Inner\_constraints:= Inner\_constraints\cup C\_inner$\;
$F_{max}:= F_{max}\wedge C_{width}\wedge C_{height}$\;
{\tcc{reflect the intrinsic impact of $F_{sub}(w)$ to the outer layer}}
}
$C_{out\_width}:=$ soft\_constraints\_hardending($F_{max}$,$[screen]_w$)\;
$Outer\_constraint:= F_{max}.hard\wedge C_{out\_width}${\tcc{convert $F_{max}$ to outer-layer SMT formula}}
{\bf Return} ($Outer\_constraint$, $Inner\_constraints$)\;
\end{algorithm}
 The $sub\_constaints$ of independent widgets can be extracted from the original formula to solve independently.
After the extraction, the original formula is divided into two layers: the {\bf outer} layer illustrates the overall layout, composed of the original constraints without the $sub\_constaints$ of independent widgets; the {\bf inner} layer consists of multiple formulas specifying the layout of independent widgets based on their $sub\_constraints$.
Such extraction can be viewed as omitting the internal details of the sub-layouts.

However, directly extracting $sub\_constraints$ from the original formulas can lose the embedded intrinsic constraints on independent widgets.
For example, $F_{sub}(Wide\_bar)$ limits the lower bound of $[Wide\_bar]_w$, and simply extracting it from the outer layer will omit such constraints.
Without such intrinsic constraints, $[Wide\_bar]_w$ can be assigned with a smaller value than its lower bound by solving the outer layer formulas, but the $F_{sub}(Wide\_bar)$ will become unsatisfiable accordingly.

To avoid losing the intrinsic information embedded in the extracted constraints, the influence of $sub\_constraints$ on its corresponding independent widget should be detected and reflected in the outer-layer formula.
We apply the {\it Interval-based soft clauses hardening} preprocessing method to the extracted $sub\_constraints$ of independent widgets, detecting the intrinsic constraints of $sub\_constraints$ on their size properties.
These detected constraints are integrated into the outer-layer formula to reflect the impact of $sub\_constraints$.

Note that, to avoid mutual influence between width and height, which complicates the range partitioning, we only choose independent widgets with irrelevant height and width properties.
Moreover, to make the independent widgets efficient to solve, the number of clauses in the extracted $sub\_constraints$ of each independent widget is limited in the threshold $t=300$.

Alg.~\ref{inner_outer} describes the {\it Independent widgets extraction} preprocessing method in detail.
Specifically, we first extract the $sub\_cosntraints$ of independent widgets from the original MaxSMT formula (Line 2).
For each independent widget $w$, the intrinsic constraint of $F_{sub}(w)$ on its width and height properties is detected as $C_{width}$ and $C_{height}$ by applying the {\it Interval-based soft constraints hardening} method (Lines 4--5).
Then the $sub\_cosntraints$ is converted to an inner-layer SMT formula $C_{inner}$ (Line 6), and inserted into the inner-layer constraint set $Inner\_constraints$ (Line 7).
The intrinsic influence of $sub\_cosntraints$ on the independent widget is reflected in the outer layer (Line 8).
Finally, the {\it Interval-based soft constraints hardening} method is applied to the MaxSMT formula in the outer layer, converting it into an SMT formula $C_{out}$ (Lines 9--10).
The preprocessing method returns the overall constraints in the outer layer and the set of constraints specifying the sub-layout of independent widgets in the inner layer, both of which are converted into SMT formulas (Line 11).

\begin{example}
As shown in Fig.~\ref{extract}, $Wide\_bar$ and $3\_col\_Table$ are independent widgets, and thus their $sub\_constraints$ can be extracted from the original formulas.
As shown in Fig.~\ref{outer}, the internal details of independent widgets are omitted in the outer layer.
After the extraction, the intrinsic influence on their size property is detected and reflected in the outer layer formula.
\end{example}

Note that the influence of $sub\_constraints$ on other properties can be omitted for the following reasons:
the $visibility$ property can be dismissed because the corresponding $sub\_constraints$ of those invisible independent widgets do not need to be solved.
Moreover, since the sub-layout can be translated, the positional properties of the independent widget $w$, namely $[w]_x$ and $[w]_y$, can be simplified as 0 in the extracted constraints.
After determining the sub-layout by solving the corresponding inner-layer formula, the position of all widgets in the sub-layout can shift according to $[w]_x$ and $[w]_y$.

\section{Preview module} 
\label{preview module}
Before deploying these SMT formulas to the terminal end, it is essential to solve the converted formulas on the development end and preview the results to ensure that the layout meets the expectations.
Moreover, the following factors may result in potential issues, which can be addressed in advance:

First, the original MaxSMT formula specified by the designer may contain conflict hard clauses, which only take effect at certain sizes.
In this scenario, the ``unsatisfiable core'' function of SMT solvers should be employed to identify the conflict constraint sets and notify the designers.

Secondly, the predetermined truth assignment of the soft constraints may become infeasible at certain sizes.
Specifically, given the truth assignments of soft constraints, denoted as $\alpha_{soft}$, the {\it Interval-based soft constraints hardening} preprocess identifies the corresponding intervals by utilizing the OMT solver to determine their maximum and minimum values.
This is based on the assumption that the interval is continuous and that $\alpha_{soft}$ is feasible within the interval.
However, although the probability is low, the feasible interval corresponding to $\alpha_{soft}$ values may be discontinuous,
and the SMT formula becomes unsatisfiable when the size property $p$ takes the value $curr\_p$ within these gaps.

In the second situation, the {\it Interval-based soft constraints hardening} preprocess should be reapplied to determine the continuous feasible interval for $\alpha_{soft}$.
Specifically, it should be applied to the MaxSMT formula attached with the constraint $(\bigwedge_{\ell\in \alpha_{soft}}\ell)\rightarrow (p>curr\_p)$, ensuring that $(p>curr\_p)$ is enforced when determining the lower bound corresponding to $\alpha_{soft}$.
Detailed descriptions of the preview process algorithm can be found in Appendix \ref{appendix_a}.

\begin{example}
Suppose the discontinuous feasible interval corresponding to $\alpha_{soft}$ is $[1000,1400]\cup[1600,2000]$.
The {\it Interval-based soft constraints hardening} preprocess determines the minimum and maximum values of the size property $p$ as 1000 and 2000 respectively, and $[1000,2000]$ is set as the corresponding interval.
However, the SMT formula becomes unsatisfiable at the size value $p=1500$.
To find the continuous feasible interval, the preprocess is reapplied to the MaxSMT formula attached with $(\bigwedge_{\ell\in \alpha_{soft}}\ell)\rightarrow (p>1500)$.
When determining the lower bound for $\alpha_{soft}$, the constraint $(p>1500)$ takes effect.
Consequently, the continuous interval $[1600,2000]$ can be determined.
\end{example}

\section{Solving module}
After the preprocessing and previewing module, the simplified constraints are deployed to the terminal end.
A two-level solving strategy is proposed to solve the pre-processed formulas according to the current screen size, leveraging the reasoning and incremental solving power of SMT solvers.
Based on the solving strategy, two SMT solvers are adopted as backend solvers:
\begin{itemize}
    \item A mainstream complete SMT solver named Z3~\cite{moura2008z3}
    \item A lightweight local search SMT solver named LocalSMT~\cite{cai2023local}, customized to this incremental scenario to improve the solving efficiency and decrease memory usage.
\end{itemize}

\subsection{A Two-level Solving strategy}
\label{two-level}

As the screen width varies, these SMT formulas are solved by {\it incrementally} modifying the constraints on screen width or size properties of independent widgets.

After the {\it Interval-based soft constraints hardening} preprocessing procedure, the range of size properties is divided into multiple intervals, and the corresponding truth assignments of soft constraints are determined in each interval.
Although these Boolean assignments can be inferred from the constraints, we explicitly evaluate them according to the interval in which the current size property is located.
These assignments are incorporated into the SMT formulas as unit clauses, prompting the SMT solvers to apply {\it Boolean automated reasoning} based on the Boolean constraints of hierarchical information.

With the {\it reasoning} power of SMT solvers, invisible containers can be identified and their related positional constraints can be excluded in advance, drastically reducing the formula scale.
The truth assignments of soft constraints are updated {\it incrementally} when the current size property enters a new interval.
Since the assignments of soft constraints remain constant within an interval, the Boolean reasoning can be reused within the interval, and there is no need to redo it every time the size properties vary.

{\bf Abstraction and Refinement: }
After the {\it Independent widgets extraction} preprocessing method, the constraints are divided into two layers: the outer layer contains an SMT formula that illustrates the overall layout, while the inner layer consists of multiple SMT formulas specifying the sub-layouts of independent widgets.
The solving process is divided into two levels: {\it Abstraction} and {\it Refinement}, targeting the outer and inner layer formulas respectively.
In the {\it abstraction} level, the outer layer constraints are solved, and the properties of independent widgets are determined.
In the {\it refinement} level, the sub-layouts of those visible independent widgets are determined by solving their corresponding $sub\_constraints$, attached with their size properties.
Then these sub-layouts are translated according to the corresponding position properties of independent widgets.

\begin{example}
In the {\it abstraction} level shown in Fig.~\ref{outer}, the overall layout is determined by solving the outer-layer constraints, according to the assignment of soft constraints regarding the current screen width.
Then in the {\it refinement} level shown in Fig.~\ref{refine}, the size properties of the visible independent widgets, $Wide\_bar$ and $3\_col\_Table$, are transmitted to the inner-layer constraints, and their sub-layouts are determined and relocated according to the position properties.
\end{example}

\subsection{Customize LocalSMT as Backend Solver}
\label{sls for layout}
In addition to a mainstream complete SMT solver called Z3~\cite{moura2008z3}, 
a lightweight local search SMT solver called LocalSMT is customized as the backend solver for the following reasons:

First, most constraints remain satisfied as the screen width varies.
In such an incremental scenario, LocalSMT can utilize local search to incrementally modify the current solution and efficiently find a new solution near the current one.

Second, LocalSMT is a lightweight local search solver dedicated to SMT formulas with the arithmetic theory and has a compact data structure.
Thus, it can save memory usage and is more suitable for the terminal end with rigorous memory restrictions.
In contrast, complete SMT solvers with general purpose occupy much memory, as they typically support abundant theories and maintain complex context information.

Nevertheless, after applying Boolean automated reasoning, the SMT formulas still contain numerous equations illustrating the positional relationship among widgets. 
LocalSMT can get stuck on these equations because its "critical move" operator cannot comprehensively investigate the intrinsic relationships among them.
Therefore, a simplification method called {\bf Unit equation elimination} is proposed to customize LocalSMT for layout constraints.
This method explores the relationship among equations and eliminates redundant variables.
The detailed description is as follows:

Given an SMT formula, we first extract the equations in unit clauses, denoted as {\it unit equations}, as a system of linear equations.
Then the Guass-Jordan elimination~\footnote{\url{https://en.wikipedia.org/wiki/Gaussian_elimination}} is employed to resolve these linear equations.
If there are solutions in the form of $x=k$ or $y=k*x+b$, where $x$ and $y$ are variables, and $k$ and $b$ are parameters, these variables can be eliminated by substituting them with the corresponding expressions.

After the elimination, additional {\it unit equations} may be introduced for the following reasons:
first, once the truth values of certain constraints are established by preset variables, new unit clauses may be introduced.
Secondly, certain inequalities will combine with other inequalities to generate new equations after the elimination.
A typical situation is illustrated in the following example:

\begin{example}
    The variable $z$ is set to 0 after the {\it Unit equation elimination}.
    The clause $(x+y=0)\vee(z>0)$ is converted to a {\it unit equation}, as $z>0$ has already been established as false.
    The clauses $(a+b+z\ge0)$ and $(a+b\le0)$ are converted to a new {\it unit equation} $(a+b=0)$.
\end{example}

Thus, the {\it Unit equation elimination} procedure outlined above should be repeated until no more variables can be eliminated.


\section{Evaluation}
\label{evaluation}
Based on the proposed strategies, we implement SMT-Layout and evaluate it on multiple benchmarks derived from real-world layouts.
These specifications are deployed to two terminal devices to evaluate the performance: a laptop with an Intel i7 CPU, and a mobile phone with a Snapdragon 8 Gen 2 CPU.
The solving efficiency is crucial to facilitate the real-time interaction of real-world layouts as the screen width varies.
Moreover, memory usage is also a critical metric for measurement, since the terminal equipment typically has rigorous memory usage restrictions.
By traversing the range of screen width for 10 runs, we measure the average interaction time, maximum interaction time, and memory usage.
Note that the run time is measured in milliseconds while memory usage is measured in MB.
To save space, in the following context, the interaction time is presented in the form of Average (Maximum).

\begin{figure}[]
    \centering
    \begin{subfigure}[b]{0.28\linewidth}
      \includegraphics[height=7cm]{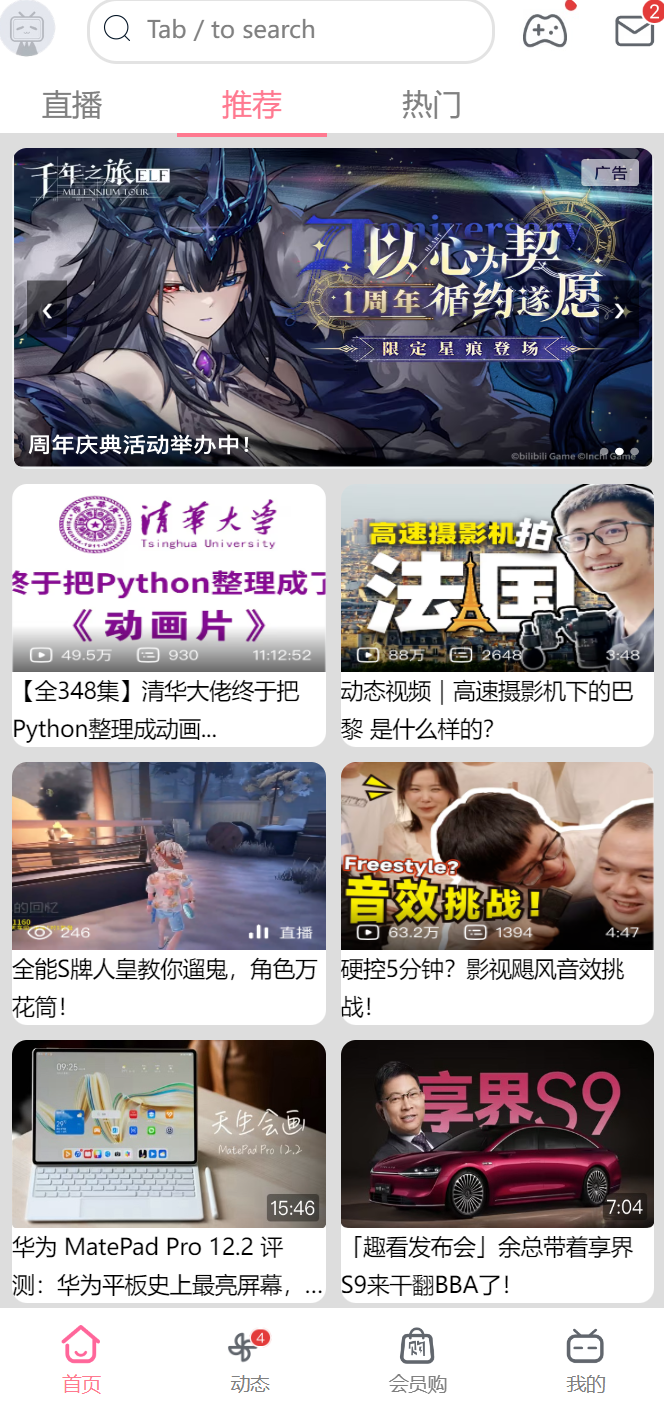}
      \caption{Mobile phone version}
    \end{subfigure}
    \begin{subfigure}[b]{0.7\linewidth}
      \includegraphics[height=7cm]{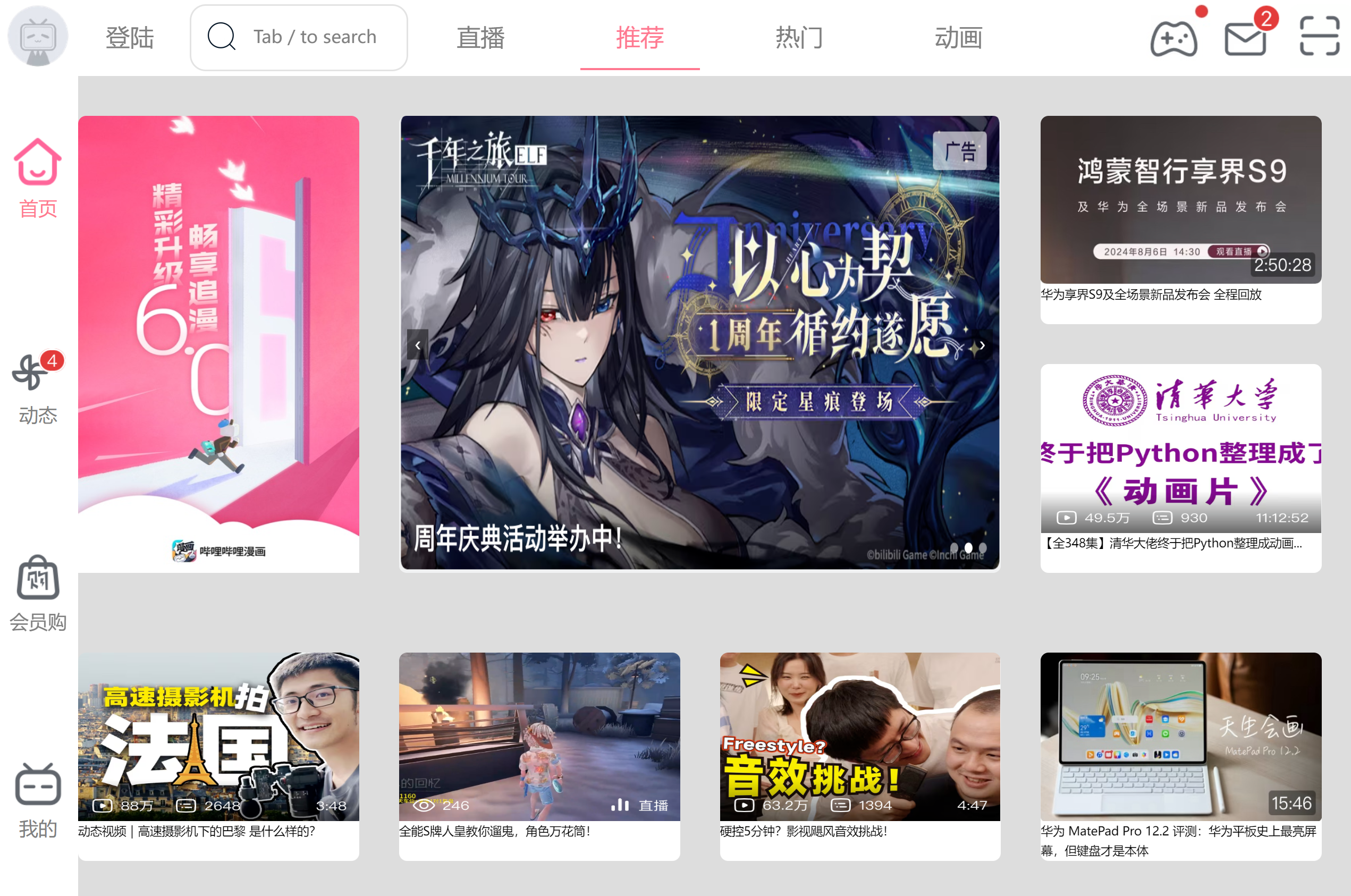}
      \caption{TV version }
    \end{subfigure}

    \begin{subfigure}[b]{0.45\linewidth}
      \includegraphics[height=7cm]{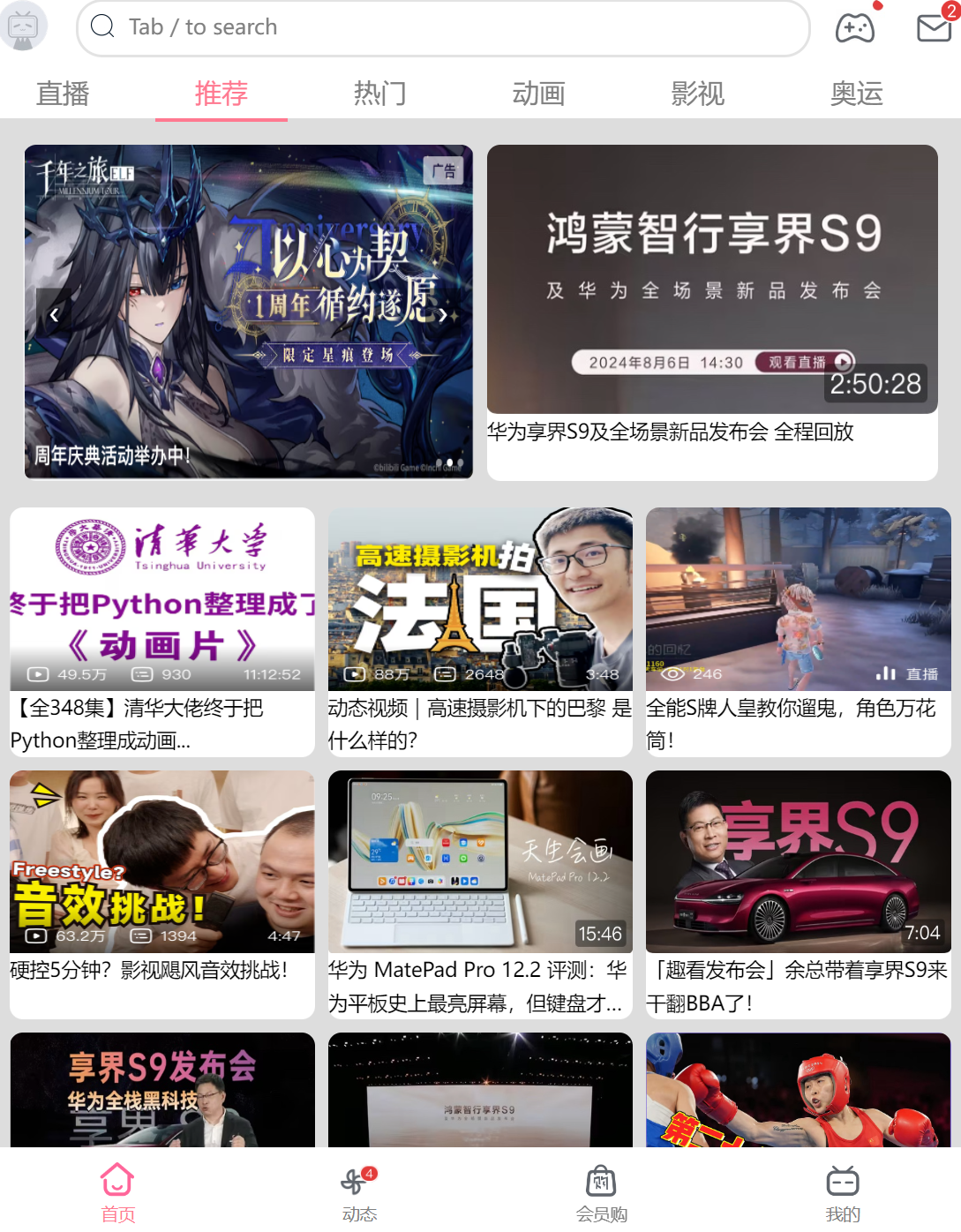}
      \caption{Tablet version}
    \end{subfigure}
    \begin{subfigure}[b]{0.5\linewidth}
      \includegraphics[height=7cm]{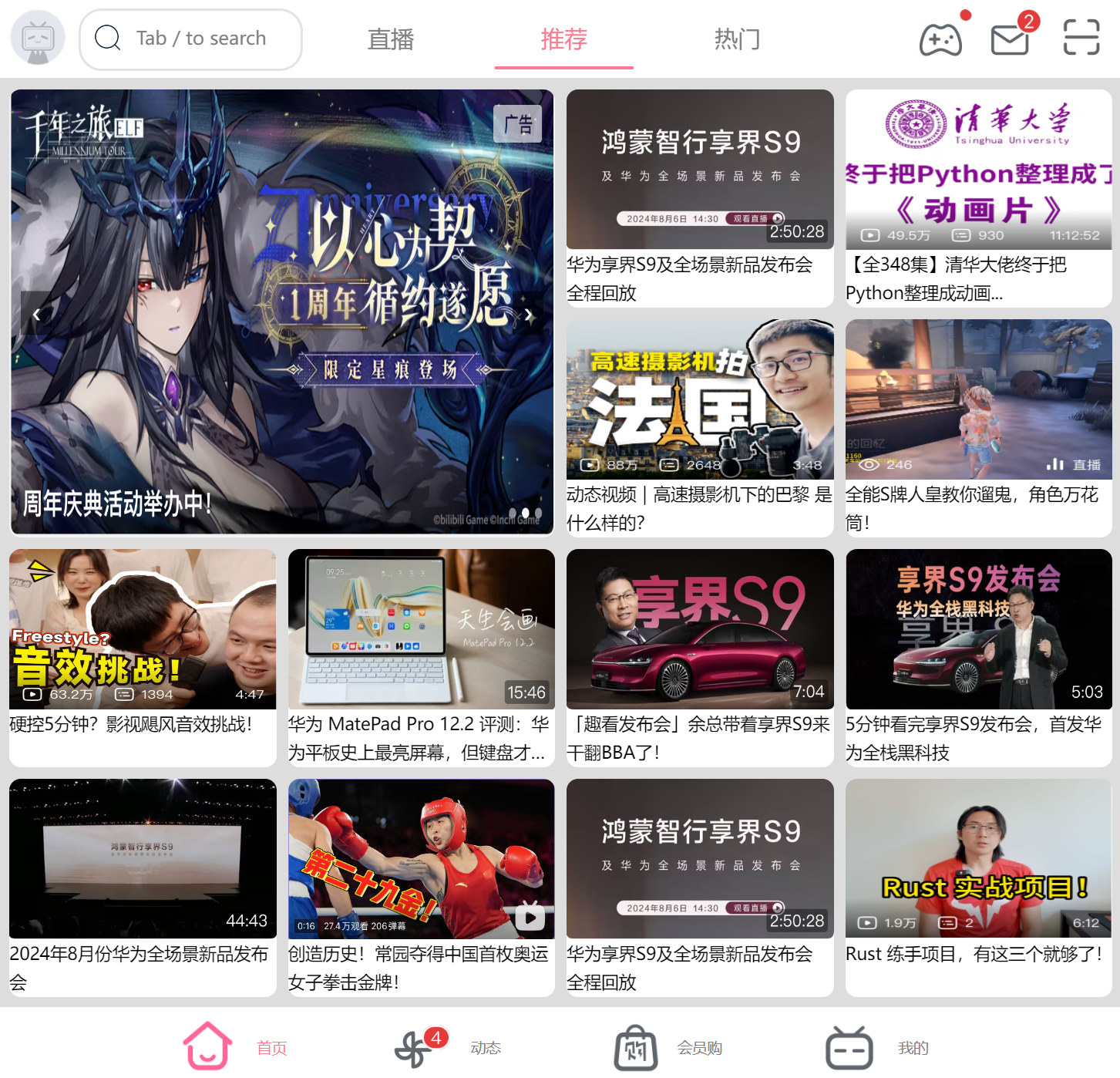}
      \caption{PC version}
    \end{subfigure}
    \caption{Different layouts of a mobile App called Bilibili are presented as the screen width varies: namely (a) the mobile phone version, (c) the Tablet version, (d) the PC version, and (b) the TV version. These real-world GUI layouts can adapt to multiple screen sizes with only one specification and support real-time interaction when resizing the screen.}
    \label{fig_bilibili}
\end{figure}

{\bf Comparison declaration:} Note that SMT-Layout is not compared with ORCSolver~\cite{jiang2019orc,jiang2020orcsolver}, which can also support various screen sizes with only one specification.
The reasons are as follows:

First, SMT-Layout is dedicated to real-world layouts, and the benchmarks are derived from actual GUIs.
However, ORCLayout mainly focuses on the Flow layout pattern and does not support abundant layout containers to specify these real-world layouts.
Specifically, We have consulted the authors of ORCLayout, and they admit that ORCLayout is unsuitable for practical layout design because the widgets in ORCLayout are compactly adjacent.

Second, ORCSolver has internal bugs and frequently crashes during solving, preventing it from being executed in comparison experiments.
Specifically, we have downloaded the source code from their repository~\footnote{https://github.com/YueJiang-nj/ORCSolver-CHI2020}, and we find that ORCSolver crashes even on almost all original examples~\footnote{We tried all original examples in the directory ``/ORCSolver-CHI2020/Code/ORCSolver (Ours)''. We find that all examples encounter the error ``std::out\_of\_range'' when calling the embedded CVXPY solver, with only one exception ``flow\_around\_pattern.py''.}.

As a reference, according to the results reported in their paper, the solving time of ORCsolver on benchmarks of a similar scale is several hundred milliseconds, which is considerably slower than our approach.
Even on the benchmark of the smallest scale, consisting of only 20 widgets, the average reported interaction time reaches up to 149 milliseconds, unable to meet the requirement for real-time interaction.

\subsection{Implementation}
The input module is implemented in C++, and the C++ API is provided for designers to specify layouts.
The preprocessing and previewing modules are implemented in Python, and the Python API of Z3~\cite{moura2008z3} is utilized to solve the (Max)SMT formulas.
After deploying the SMT formulas to the terminal end, the solving module is implemented in C++ to enhance efficiency.
Two SMT solvers are employed as the alternative backend solvers in the solving module: the C++ API of Z3 and the LocalSMT solver customized for layout constraints.
The solving module is compiled by Emscripten~\cite{zakai2011emscripten}, and the resulting layout is displayed on the browser.

\subsection{Benchmarks}
Using our C++ API, we specified 12 benchmarks derived from real-world GUI layouts, including web pages and mobile App GUIs.
These instances cover various typical GUI application scenarios, including e-commerce websites, streaming platforms, portal websites, development platforms, etc.
These benchmarks represent typical layout page sizes, consisting of 113 widgets and 2229 constraints on average.
Each of these instances can adjust to various screen sizes with only one specification.

For example, Fig.~\ref{fig_bilibili} illustrates a typical benchmark named Bilibili\_app: different layouts of a mobile App called Bilibili are displayed on equipment with varying screen widths, including mobile phones, Tablets, PCs, and TVs. 
These different GUI layouts can adapt to various screen sizes with only one specification, and real-time interaction is required as the screen width varies.

Table. \ref{benchmark} describes the detailed information of these benchmarks.
The $url$ denotes the address of the real-world web pages.
Since Bilibili\_app is the GUI of a mobile App, it does not have a $url$ address.
The number of widgets, variables, hard clauses in the original MaxSMT formula, Placeholder containers, and soft clauses are denoted as \#(W), \#(V), \#(hard\_c), \#(P), and \#(soft\_c), respectively.
Note that there may be a situation where \#(V)$>5*\#$(W) because certain containers can introduce auxiliary variables.

These benchmarks and their corresponding demos deployed at the terminal end can be found in the anonymous repository~\footnote{\url{https://anonymous.4open.science/r/SMT-Layout/}}.
The real-time interaction on these layouts is also presented in the supplementary video.

\begin{table}[]
\caption{layout benchmark information}
\label{benchmark}
\begin{tabular}{@{}lllllll@{}}
\toprule
Name                &url                          & \#(W) & \#(V) & \#(hard\_c) & \#(P) & \#(soft\_c) \\ \midrule
Meituan             &\url{www.meituan.com}        & 101         & 505           & 1789       & 4                & 8                 \\
Software            &\url{www.is.cas.cn}          & 114         & 574           & 2166       & 4                & 8                 \\
Bilibili\_web            &\url{www.bilibili.com}       & 95          & 475      & 1676       & 1                & 2                 \\
Jingdong            &\url{www.jd.com}             & 102         & 510           & 1495       & 1                & 2                 \\
Amazon              &\url{www.amazon.com}         & 50          & 250           & 865       & 1                & 2                 \\
Amazon\_women       &\url{www.amazon.com/women}   & 73          & 365           & 1168       & 1                & 1                 \\
Amazon\_video       &\url{www.amazon.com/video}   & 80          & 400           & 1282       & 2                & 5                 \\
Amazon\_outlet      &\url{www.amazon.com/outlet}  & 121         & 605           & 3689       & 1                & 4                 \\
Github              &\url{www.github.com}             & 84          & 420       & 1404       & 4                & 8                 \\
Github\_Z3          &\url{www.github.com/Z3Prover/Z3} & 147         & 735       & 2951       & 4                & 8                 \\
App\_store          &\url{appgallery.huawei.com}  & 207          & 1035           & 4175       & 7                & 17                \\
Bilibili\_app       &                             & 178         & 893           & 4092       & 3                & 11                \\
\bottomrule
\end{tabular}
\end{table}

\subsection{Overall results}
\begin{table}[]
\caption{Overall results of SMT-Layout with Z3 and LocalSMT serving as the backend SMT solvers. }
\label{overall_result}
\begin{tabular}{@{}llllllll@{}}
\toprule
benchamrk      & \multicolumn{3}{c}{SMT-Layout(Z3)}                      &  & \multicolumn{3}{c}{SMT-Layout(LocalSMT)}                \\ \cmidrule(lr){2-4} \cmidrule(l){6-8} 
               & PC\_Time              & Phone\_Time              & Memory     &  & PC\_Time               & Phone\_Time              & Memory \\ \midrule
Meituan        & 4.2 (8.5)          & 7.1  (12.8)           & 62.4       &  & 0.19 (1.7)          & 0.21 (2.5)            & 2.8        \\
Software       & 3.9 (9.7)          & 6.6  (14.6)           & 44.2       &  & 0.12 (1.1)          & 0.15 (1.6)            & 2.9        \\
Bilibili\_web  & 3.6 (5.8)          & 5.8  (8.4)            & 17.9       &  & 0.13 (0.34)         & 0.17 (0.48)           & 2.1        \\
Jingdong       & 2.9 (4.8)          & 5.4  (7.3)            & 31.2       &  & 0.13 (0.35)         & 0.16 (0.51)           & 1.9        \\
Amazon         & 1.4 (2.6)          & 2.6  (3.9)            & 39.2       &  & 0.07 (0.25)         & 0.09 (0.48)           & 1.3        \\
Amazon\_women  & 3.5 (5.0)          & 6.7  (7.8)            & 27.4       &  & 0.22 (3.07)         & 0.32 (4.4)            & 1.7        \\
Amazon\_video  & 2.1 (3.7)          & 3.6  (5.6)            & 33.4       &  & 0.09 (0.55)         & 0.12 (0.89)           & 1.8        \\
Amazon\_outlet & 10.4 (12.5)        & 12.9 (15.1)           & 24.3       &  & 0.26 (5.6)          & 0.37 (7.8)            & 4.2        \\
Github         & 1.9 (2.3)          & 3.3  (4.3)            & 46.2       &  & 0.18 (0.7)          & 0.21 (1.2)            & 2.2        \\
Github\_z3     & 3.6 (6.3)          & 6.6  (9.8)            & 72.7       &  & 0.81 (7.5)          & 0.96 (10.6)           & 4.7        \\ 
App\_store     & 7.6 (14.2)         & 13.8 (21.8)           & 112.7      &  & 0.25 (5.8)          & 0.37 (8.2)            & 5.3        \\
Bilibili\_app  & 7.8 (14.8)         & 12.1 (22.5)           & 35.9       &  & 0.95 (10.1)         & 1.4 (13.4)            & 3.9        \\ \bottomrule
\end{tabular}
\end{table}

Table.~\ref{overall_result} presents the overall results for 2 versions of SMT-Layout that respectively adopt Z3 and LocalSMT as backend solvers, denoted as SMT-Layout(Z3) and SMT-Layout(LocalSMT).
Note that, since the memory usage on the PC and the mobile phone are similar, we only report the memory usage on the PC to save space.

{\bf Efficiency: }The result indicates that the solving module can support real-time interaction with both backend SMT solvers.
Specifically, the average and maximum interaction time on each benchmark is less than 25 ms, which is the frame time at a 40Hz refresh rate, even on equipment with limited computational power, such as a mobile phone.
Moreover, when applying LocalSMT as the backend solver, the responsive speed is significantly faster than the counterpart adopting Z3.

{\bf Memory usage: }
The result confirms that LocalSMT occupies minimal memory due to its compact data structure, and thus it is appropriate for equipment with rigorous memory limitation.
In contrast, when applying Z3, a significant amount of memory is required because the {\it Independent widget extraction} preprocessing method can generate multiple SMT formulas, and for each formula, a Z3 solver instance that maintains complex context information should be reserved.

\subsection{Effectiveness of Proposed strategies}
Three configurations are proposed to evaluate the effectiveness of the preprocessing methods.
Note that the {\it Interval-based soft constraints hardening} and {\it Independent widgets extraction} preprocessing methods are denoted as $Hard$ and $Ind$ for short.
\begin{itemize}
    \item $\nu Z$: the baseline version that directly solves the original MaxSMT formulas with the MaxSMT solver $\nu Z$~\cite{bjorner2015nuz}, the default optimization algorithm of Z3. This is also the solving method adopted in~\cite{jiang2019orc} to solve ORCLayout.
    \item {\it Hard+Z3}: the modified version that only applies the {\it Interval-based soft constraints hardening} method as the preprocessing method, with Z3 serving as the backend SMT solver.
    \item {\it Hard+LocalSMT}: the modified version that only applies the {\it Interval-based soft constraints hardening} preprocessing method, with LocalSMT as the backend SMT solver.
\end{itemize}
To make the comparison more intuitive, SMT-Layout(Z3) and SMT-Layout(LocalSMT) are denoted as {\it Hard+Ind+Z3} and {\it Hard+Ind+LocalSMT} in this context, indicating that both preprocessing methods are adopted.
Since the solving time on the mobile phone is almost always about 1.5 times that on PCs, the subsequent evaluations are only conducted on the PC.
The results are presented in Table. \ref{effect}, and the versions with the shortest responsive time and lowest memory usage are emphasized in {\bf bold}.

Moreover, we compare the detailed information of SMT formulas after applying the {\it Boolean reasoning} and the {\it Unit equation elimination} method, to confirm the effectiveness of these simplification methods.
Specifically, in Table. \ref{clauses_num}, we present the number of clauses, non-unit clauses, and unit equations in the original SMT formulas and the counterparts after these simplifications, denoted as \#(C), \#(NU), and \#(UE) for short.

\begin{sidewaystable}
\caption{Results of different configurations}
\label{effect}

\begin{tabular}{@{}lllllllllllllll@{}}
\toprule
benchmark      & \multicolumn{2}{c}{$\nu$Z} &  & \multicolumn{2}{c}{{\it Hard+Z3}} & \multicolumn{1}{c}{} & \multicolumn{2}{c}{{\it Hard+Ind+Z3}} & \multicolumn{1}{c}{} & \multicolumn{2}{c}{{\it Hard+LocalSMT}} & \multicolumn{1}{c}{} & \multicolumn{2}{c}{{\it Hard+Ind+LocalSMT}} \\ \cmidrule(lr){2-3} \cmidrule(lr){5-6} \cmidrule(lr){8-9} \cmidrule(lr){11-12} \cmidrule(l){14-15} 
               & Time    & Memory    &  & Time       & Memory       &         & Time     & Memory    &         & Time        & Memory      &          & Time        & Memory       \\ \midrule
Meituan        & 54.6 (79.8)      & 5.7        &  & 5.5 (12.6)        & 8.1           &         & 4.2 (8.5)       & 62.4       &         & 0.34 (5.2)         & 3.5          &          & {\bf  0.19 (1.7)   }         & {\bf 2.8}           \\
Software       & 86.9 (107.7)      & 6.2        &  & 4.8 (14.5)        & 9.2           &         & 3.9 (9.7)       & 44.2       &         & 0.39 (6.4)         & 3.8          &          & {\bf  0.12 (1.1)   }         & {\bf 2.9}           \\
Bilibili\_web  & 87.4 (104.8)      & 4.1        &  & 4.6 (6.8)         & 8.2           &         & 3.6 (5.8)       & 17.9       &         & 0.21 (0.37)        & 2.4          &          & {\bf  0.13 (0.34)  }         & {\bf 2.1}           \\
Jingdong       & 51.9 (71.4)      & 4.6        &  & 3.5 (5.7)         & 8.0           &         & 2.9 (4.8)       & 31.2       &         & 0.23 (0.62)        & 2.4          &          & {\bf  0.13 (0.35)  }         & {\bf 1.9}           \\
Amazon         & 38.3 (52.3)      & 3.5        &  & 2.4 (2.8)         & 5.3           &         & 1.4 (2.6)       & 39.2       &         & 0.12 (0.34)        & {\bf 1.2}    &          & {\bf  0.07 (0.25)  }         & 1.3           \\
Amazon\_women  & 55.3 (77.3)      & 4.7        &  & 4.1 (8.2)         & 7.3           &         & 3.5 (5.0)       & 27.4       &         & 0.28 (4.2)         & {\bf 1.5}    &          & {\bf  0.22 (3.07)  }         & 1.7           \\
Amazon\_video  & 52.4 (87.5)      & 6.5        &  & 3.6 (4.8)         & 7.6           &         & 2.1 (3.7)       & 33.4       &         & 0.21 (2.8)         & 2.3          &          & {\bf  0.09 (0.55)  }         & {\bf 1.8}           \\
Amazon\_outlet & 73.9 (92.4)      & 7.6        &  & 14.2 (21.6)       & 11.7          &         & 10.4 (12.5)     & 24.3       &         & 0.29 (7.2)         & 4.5          &          & {\bf  0.26 (5.6)   }         & {\bf 4.2}           \\
Github         & 59.5 (71.6)      & 8.3        &  & 2.6 (5.5)         & 6.9           &         & 1.9 (2.3)       & 46.2       &         & 0.48 (2.7)         & 3.0          &          & {\bf  0.18 (0.7)   }         & {\bf 2.2}           \\
Github\_z3     & 73.9 (124.2)     & 12.6       &  & 4.2 (8.6)         & 9.6           &         & 3.6 (6.3)       & 72.7       &         & 1.25 (14.3)        & {\bf 4.3}    &          & {\bf  0.81 (7.5)   }         & 4.7           \\
App\_store     & 272.4 (335.6)    & 17.5       &  & 10.2 (20.4)       & 11.8          &         & 7.6 (14.2)      & 112.7      &         & 3.1 (24.3)         & 13.6         &          & {\bf  0.25 (5.8)   }         & {\bf 5.3}           \\
Bilibili\_app  & 153.5 (334.5)    & 14.4       &  & 8.4 (16.2)        & 10.4          &         & 7.8 (14.8)      & 35.9       &         & 1.05 (10.6)         & 4.2          &          & {\bf 0.95 (10.1)  )}         & {\bf 3.9}           \\ \bottomrule
\end{tabular}
\end{sidewaystable}

\begin{table}[]
\caption{Detailed clause information: the number of clauses, non-unit clauses, and unit equations in original SMT formulas, the SMT formulas after Boolean automated reasoning, and the SMT formulas after {\it Unit equation elimination} method.}
\label{clauses_num}
\begin{tabular}{@{}llllllllllll@{}}
\toprule
\multirow{2}{*}{benchmark} & \multicolumn{3}{c}{Original} &  & \multicolumn{3}{c}{Boolean Reasoning}  &  & \multicolumn{3}{c}{Unit\_Eq\_Elimination} \\ \cmidrule(lr){2-4} \cmidrule(lr){6-8} \cmidrule(l){10-12} 
                           & \#(C) & \#(NU) & \#(UE) &  & \#(C) & \#(NU) & \#(UE) &  & \#(C)      & \#(NU)     & \#(UE)     \\ \midrule
Meituan                     & 1847 & 1101 & 62  &  & 1083 & 4   & 327 &  & 133  & 4  & 4            \\
Software                    & 2208 & 1308 & 62  &  & 1449 & 8   & 440 &  & 131  & 4  & 1            \\
Bilibili\_web               & 1689 & 652  & 247 &  & 963  & 4   & 398 &  & 93   & 4  & 2            \\
Jingdong                    & 1503 & 398  & 291 &  & 1026 & 4   & 404 &  & 135  & 4  & 15           \\
Amazon                      & 873  & 430  & 64  &  & 582  & 4   & 192 &  & 72   & 4  & 4            \\
Amazon\_women               & 1188 & 91   & 295 &  & 734  & 4   & 319 &  & 159  & 4  & 17           \\
Amazon\_video               & 1304 & 322  & 274 &  & 693  & 0   & 309 &  & 95   & 0  & 2            \\
Amazon\_outlet              & 3709 & 2548 & 302 &  & 1226 & 4   & 546 &  & 102  & 4  & 14           \\
Github                      & 1429 & 798  & 99  &  & 933  & 8   & 305 &  & 149  & 8  & 3            \\
Github\_z3                  & 3018 & 1992 & 126 &  & 1877 & 156 & 516 &  & 132  & 4  & 2            \\
App\_store                  & 4267 & 2462 & 189 &  & 2404 & 784 & 459 &  & 1072 & 0  & 58           \\
Bilibili\_app               & 4159 & 2796 & 81  &  & 1537 & 48  & 389 &  & 406  & 48 & 0          \\ \bottomrule
\end{tabular}
\end{table}

{\bf Interval-based soft constraints hardening: }
The result of $\nu Z$ and {\it Hard+Z3} in Table \ref{effect} confirms that the {\it soft constraint hardening} preprocessing method can considerably enhance the solving efficiency, reducing the average(maximum) interaction time by 93.6\%(91.6\%).
The primary reason is that after determining the truth value of soft constraints, the SMT solvers can infer the $visibility$ properties of other widgets by Boolean automated reasoning, thereby omitting the constraints of invisible containers.
Specifically, as shown in the {\it Original} and {\it Boolean Reasoning} column of Table. \ref{clauses_num}, after determining the truth assignment of soft constraints and applying the Boolean automated reasoning, 46.3\% of clauses are eliminated, significantly reducing the search space.
Furthermore, after the Boolean automated reasoning, most clauses are converted to unit clauses, indicating that the potential combinations are substantially reduced and the SMT formulas can be efficiently solved.

{\bf Independent Widgets Extraction:}
comparing the result of {\it Hard+Z3} and {\it Hard+Ind+Z3} in Table. \ref{effect},  the average(maximum) interaction time is reduced by 22.3\%(28.2\%) after the {\it Independent widgets extraction} preprocessing method when Z3 is employed as the backend SMT solver.
However, the memory usage increases by 4.3 times, because multiple Z3 solver instances should be reserved for all SMT formulas in the inner layer, and each instance maintains a complex context data structure.
Therefore, in the terminal end with a rigorous memory utilization limitation, it is advisable to employ the {\it Interval-based soft constraints hardening} preprocessing method exclusively when employing Z3 as the backend SMT solver.

The result of {\it Hard+LocalSMT} and {\it Hard+Ind+LocalSMT} in Table \ref{effect} confirms that the average(maximum) interaction time is reduced by 56.5\%(53.2\%) after the {\it Independent widgets extraction} preprocess when LocalSMT is employed as the backend SMT solver.
Furthermore, memory usage does not significantly rise even when multiple LocalSMT instances are reserved. 
This is due to the compact data structure of LocalSMT, where the memory usage is linearly related to the number of constraints and variables.

{\bf Unit equations elimination:}
As shown in Table \ref{clauses_num}, after applying Boolean reasoning, most clauses are converted to unit clauses, boosting the number of unit equations by 120.2\%.
Therefore, applying the {\it unit equations elimination} technique is essential to help LocalSMT investigate the intrinsic relationship among these equations.
The result shows that after this simplification, 90.1\%  of clauses can be eliminated, dramatically reducing the search space.
Furthermore, the LocalSMT solver cannot solve any layouts within 10 seconds without the {\it unit equation elimination} method, which confirms its effectiveness in preventing LocalSMT from getting stuck due to the equations.

{\bf Apply LocalSMT as Backend Solver:}
As shown in the results of {\it Hard+Ind+LocalSMT} and {\it Hard+Ind+Z3} in Table \ref{effect}, When adopting LocalSMT as the backend solver, the average(maximum) interaction time is reduced by 93.7\%(59.4\%) compared to the version adopting Z3.
This confirms that LocalSMT is significantly more efficient for this incremental scenario, where most constraints remain satisfied when the screen width varies.
The main reason is that LocalSMT can take full advantage of the current solution, and only a few variables need to be fine-tuned.
Specifically, the average local search iteration step is 15.6, confirming that only a few modifications are required to find a new feasible solution near the current one.
Moreover, the memory usage of LocalSMT is quite compact compared to its counterpart, since LocalSMT is a lightweight solver without complex context data structure, and it is dedicated to the SMT problem with arithmetic theories.

Note that when adopting LocalSMT, the maximum interaction time is 10.9 times longer than the average interaction time, and we notice that the maximum interaction time only occurs when the screen width enters a new interval divided by the $Hard$ preprocessing.
This is because the Boolean reasoning should be reapplied due to the change of soft constraints assignment, while it can be reused within the interval.

\section{Related works}

\subsection{Constraint-based layout models}
{\bf Conventional layout models} such as group, table, grid, and grid-bag layouts are widely applied to specify the layout of GUIs~\cite{myers2000past, myers1995user}.
Although these layout models are intuitive, they have the following limitations:
First,  widgets cannot be aligned across a hierarchy~\cite{lutteroth2008modular}.
Moreover, these models often implicitly add more constraints than desired, which makes them hard to maintain~\cite{zeidler2012comparing, lutteroth2006user}.
To overcome these shortcomings, more modern layout models use constraints to specify the layout, employing various constraint types and both explicit and implicit constraint specifications.

{\bf Linear constraint-based layout models } that apply linear systems are mainstream constraint-based layout models~\cite{badros2001cassowary,bill1992bricklayer,borning1997solving,hosobe2000scalable,lutteroth2008domain}:
a series of linear equations and inequalities are applied to describe the relative or absolute alignment relationships or size values of widgets.
Linear constraint-based layouts have the following advantages:
First, linear constraint systems have a well-understood mathematical foundation, which makes the results less dependent on implementation details, and thus it is easy to maintain~\cite{zeidler2012comparing,lutteroth2006user}.
Moreover, it can align widgets across hierarchies, which is impossible in conventional layout models.
Furthermore, linear layout systems can be combined to enable modular reuse of existing layout specifications~\cite{lutteroth2008modular}.
Finally, most conventional layout models can be converted to linear constraint systems~\cite{weber2010reduction,zeidler2017tiling}.

Although linear constraint-based layouts are flexible, linear constraint-based layout models share the same drawback as the conventional layout models: multiple layout alternatives need to be specified when the page size varies significantly, and the layout manager determines which layout to apply at runtime based on contexts such as screen size or aspect ratio.
This is a common approach widely applied in ``responsive web design''~\cite{natda2013responsive,mohorovivcic2013implementing,baturay2013responsive} and mobile apps~\cite{sahami2013insights,zanden1990automatic}.
These multiple layout declarations need to be synchronized, which can lead to maintenance difficulties and consistency issues between different layouts.

{\bf ORCLayout} is an adaptive constraint-based layout method based on OR constraints~\cite{jiang2019orc}, allowing the GUI layout to adapt to multiple screen sizes with only one specification.
An OR constraint is a disjunction of constraints where only one needs to be true, and soft constraints with weights are introduced to express the preference of different terms in an OR constraint. 
This can ease the burden of GUI maintenance, as designers do not need to synchronize changes between different layout specifications, making the GUI design more efficient and less error-prone.
However, ORCLayout is not suitable for real-world layout designing, because it mainly focuses on the Flow Layout patterns without abundant layout patterns supporting practical designing, and widgets are compactly adjacent in ORCLayout.

In contrast, SMT-Layout supports abundant layout containers for practical layout designing based on the flexible expression ability of (Max)SMT, and it can specify real-world GUI layouts adapting to various screen sizes using only one specification.

\subsection{Layout Solvers}
Various solvers are dedicated to the linear constraint-based specifications, using linear or quadratic programming~\cite{badros2001cassowary, bill1992bricklayer, borning1997solving, hosobe2000scalable, lutteroth2008domain}.
CLP(R) includes constraints on linear equations and inequalities in logical programming. It gradually optimizes variables but is inappropriate for interactive use~\cite{jaffar1992clp}.
DeltaBlue~\cite{freeman1990incremental,sannella1993multi} and Skyblue~\cite{sannella1994skyblue} are local propagation constraint solvers that can handle constraint hierarchies but cannot solve simultaneous constraints.
Cassowary~\cite{badros2001cassowary} employs an incremental simplex algorithm to handle linear constraints in UI layout incrementally.
QOCA~\cite{borning1997solving,marriott1998tableau,marriott2002qoca} adopted a tableau-based approach to handle constraint hierarchies.
HiRise~\cite{hosobe2000scalable} and HiRise2~\cite{hosobe2011simplex} employ a simplex method based on the LU decomposition algorithm, and an ordered constraint hierarchy is applied to solve linear layout constraints.
~\cite{MarriottMSB01} proposed an algorithm dedicated to solving disjunctions of arithmetic representing non-overlap constraints.
Note that it is based on the insight that disjuncts in a non-overlap constraint are not disjoint, while our approach handles general (Max)SMT problems without this assumption.
These solvers support soft constraints with priorities and employ penalty functions to manage them, aiming to fulfill as many soft constraints as possible according to their priority, thus improving the aesthetics~\cite{zeidler2012constraint}.

The OR constraint system can be solved by SMT solvers such as Z3~\cite{moura2008z3}, as it is essentially a special case of (Max)SMT.
However, Z3 does not perform well in this scenario because it does not fully utilize the main power of SMT solvers.
To improve the solving efficiency, ORCSolver is proposed to solve ORC Layout specifications based on a branch-and-bound (B\&B) approach with modular heuristics~\cite{jiang2020orcsolver}.
In contrast, SMT-Layout goes further and has the following advantages:

First, the ORCSolver does not support Boolean variables and cannot directly illustrate the hierarchical relationships by Boolean constraints.
Therefore, it cannot take full advantage of the automated reasoning capability of SMT solvers.
In contrast, we introduce the {\it visibility} property of widgets as Boolean variables and specify the hierarchical relationships with Boolean constraints.
Boolean automated reasoning is applied in SMT-Layout to infer the {\it visibility} of widgets, and those constraints related to invisible widgets can be omitted to reduce the search space.

Secondly, the ORCsolver does not support incremental solving and real-time interaction. In SMT-Layout, SMT
solvers are applied incrementally in our solving module, and LocalSMT is tailored for such incremental scenarios by modifying the current solution. Moreover, our solving module can achieve real-time response at the millisecond level,
even on terminal equipment with weak computational power.

Furthermore, ORC solvers employ specific modular heuristics tailored to each layout pattern, rather than relying on a general approach. New heuristics should be designed to adapt to new or more complex layout patterns. In contrast, our method is general and does not rely on specific layout patterns, allowing for the flexible enrichment of layout patterns without additional adaptation, and it can enjoy further progress in SMT solving.



\section{Conclusion}
In this paper, by fully utilizing the flexible expressive ability of (Max)SMT and the powerful solving ability of SMT solvers, we propose a novel layout approach named SMT-Layout.
SMT-Layout is the first practical constraint-based layout model that can support real-time interaction for real-world layout adapting to various screen sizes with only one specification.

SMT-Layout introduces Boolean variables to encode the hierarchical information among widgets, boosting the reasoning ability of SMT solvers.
By utilizing the abundant computing resources at the development end, two novel preprocessing methods are proposed to simplify the constraints and extract useful information to ease the solving burden at the terminal end.
After deploying the preprocessed constraints, SMT solvers, including a customized local search solver, are employed as backend solvers to solve them incrementally, leveraging their reasoning ability.
Experimental results show the efficiency of our approach, even on equipment with weak computational power and rigorous memory limitation.
Ablation experiments are conducted to confirm the effectiveness of the proposed strategies.

\subsection{Limitation \& Future work}
First, the C++ API is provided for the designers to specify the layout, which is not intuitive and difficult to acquire and understand.
Designers cannot visually specify the layouts through a GUI editor.
In our future work, we intend to develop a Domain-Specific Language (DSL) and a GUI editor to enhance the usability of our layout method.

Moreover, to employ our methods on the existing layout, designers need to manually apply APIs to specify and simulate these real-world layouts, which is often time-consuming.
Therefore, it is necessary to develop reverse engineering to simulate existing layout pages and automatically convert them to MaxSMT constraints.







\begin{appendices}

\section{Formalization of Preset Containers}
\label{preset_container}
The abundant preset containers, such as Row, Column, Flow, Waterfall, Table, and Flex Layout, are formalized in detail.
  These containers are provided for the designers to specify real-world layouts.
  The basic positional constraints in the form of (Max)SMT are automatically generated and maintained, and designers can customize them using the available APIs.
\subsection{Basic Layout Container}
A layout container $w$ contains $n$ kid widgets, denoted as $kid_i\in[w]_k$ where $i\in[1,n]$.
If the container is visible, any following constraint $c$ regarding the container takes effect: $[w]_v\rightarrow c$.
For the sake of simplicity, the fact will no longer be mentioned in the following context.

{\bf Available APIs: }
Each type of container is equipped with the subsequent fundamental APIs:
\begin{itemize}
    \item $set\_container\_width(width)$ sets the width of container $w$ as $width$: $[w]_w=width$;
    \item $set\_container\_hight(hight)$ sets the height of the container as $hight$: $[w]_h=hight$.
    \item $set\_kid\_width(width)$ sets the width of the $i$th kid widget as $width$: $[kid_i]_w=width$.
    \item $set\_kid\_hight(hight)$ sets the hight of the $i$th kid widget as $hight$: $[kid_i]_h=hight$.
\end{itemize}

\subsection{Row \& Column}
\begin{figure}[]
    \centering
    \begin{subfigure}[b]{0.45\linewidth}
    \centering
      \includegraphics[width=4.5cm]{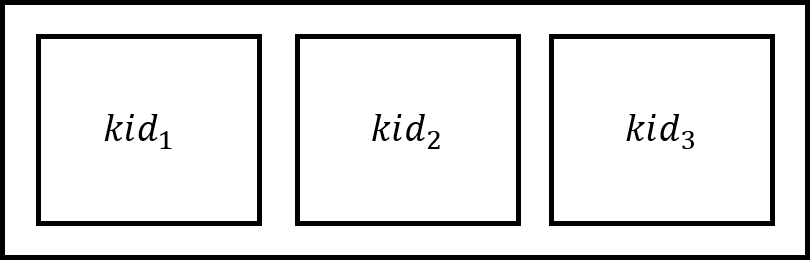}
      \caption{Row Layout Container}
    \end{subfigure}
    \begin{subfigure}[b]{0.45\linewidth}
    \centering
      \includegraphics[width=1.5cm]{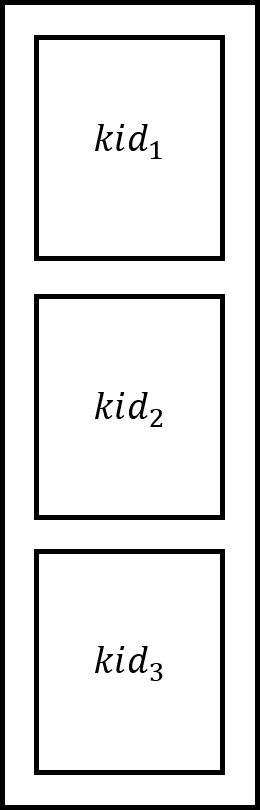}
      \caption{Column Layout Container}
    \end{subfigure}
    \caption{The Row \& Column layout container.}
    \label{fig_row_column}
\end{figure}

Row and column are basic layout containers, where all kid widgets are respectively arranged horizontally and vertically, as shown in Fig.~\ref{fig_row_column}.
Without sacrificing generality, We describe the Row container in detail, and the Column container can be analogized.

{\bf Basic Constraints:}
Given a Row container $w$ containing $n$ kid widgets $kid_i\in[w]_k$, there are 2 basic constraints:
\begin{itemize}
    \item All kid widgets are arranged in sequence without any overlapping: for each widget $kid_i,i\in[1,n-1]$, $[kid_i]_x+[kid_i]_w\le [kid_{i+1}]_x$.
    \item All kid widgets should be located within the container. Specifically, the first and last kid widgets should be horizontally enclosed within $w$: $w$: $[kid_1]_x\ge [w]_x$, $[kid_n]_x+[kid_n]_w\le[w]_x+[w]_w$; All kid widgets should be vertically contained within $w$: for each $kid_i\in [w]_k$, $[kid_i]_y\ge [w]_y$, $[kid_i]_y+[kid_i]_h\le[w]_y+[w]_h$.
\end{itemize}

{\bf Available APIs:}
$set\_margin(m,is\_equal)$ allows designers to set the margin between kid widgets. If $is\_equal$ is false, $m$ is set as the minimum value of the margin: for $i\in[2,n]$, $[kid_i]_x-[kid_{i-1}]_x-[kid_{i-1}]_w\ge m$; Otherwise, the margins are fixed as $m$: for $i\in[2,n]$, $[kid_i]_x-[kid_{i-1}]_x-[kid_{i-1}]_w=m$.

$set\_padding(p,is\_equal)$ allows designers to specify the padding to the edges. If $is\_equal$ is false, then the minimum value of the upper and bottom padding is set as $p$: for $kid_i\in [w]_k$, $[kid_i]_y-[w]_y\ge p\wedge[w]_y+[w]_h-[kid_i]_y-[kid_i]_h\ge p$, and the minimum value of the left and right padding is set as $p$: $[kid_1]_x-[w]_x\ge p\wedge[w]_x+[w]_w-[kid_n]_x-[kid_n]_w\ge p$. 
Otherwise, if $is\_equal$ is true, the corresponding inequalities are converted to the equations, indicating that the padding is fixed as $p$. 

\subsection{Flow Layout}
\label{flow_layout}
Similar to the Row and Column Container, the widgets in the Flow layout container $w$ can also be arranged horizontally and vertically. 
We only describe the horizontal version, and the vertical counterpart can be analogized.

\subsubsection{Equal-height widgets}
\begin{figure}[]
    \centering
    \begin{subfigure}[b]{0.45\linewidth}
    \centering
      \includegraphics[width=4.5cm]{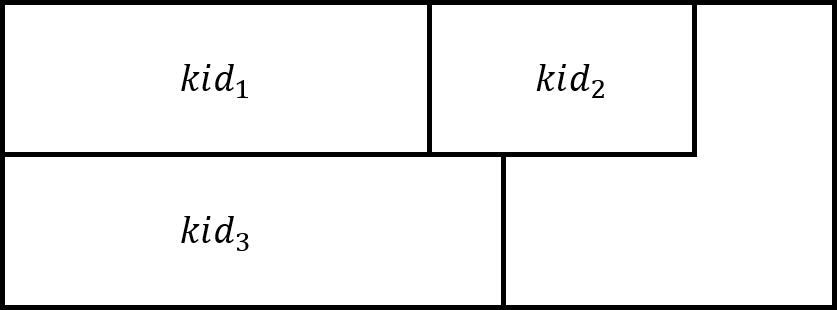}
      \caption{Wrap version}
    \end{subfigure}
    \begin{subfigure}[b]{0.45\linewidth}
    \centering
      \includegraphics[width=6.5cm]{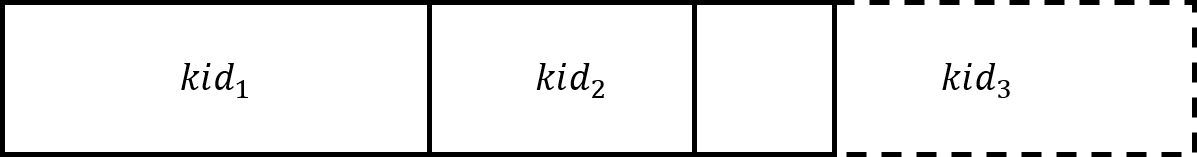}
      \caption{Non-wrap version: $kid_3$ is marked with dashed lines, indicating that it is invisible.}
      \label{non-wrap}
    \end{subfigure}
    \caption{The Flow layout with equal-height widgets.}
    \label{fig_flow_eq}
\end{figure}
For the sake of simplicity, we first assume that the heights of all widgets are identical, which is common in real-world applications: for each $kid_i\in [w]_k$ where $i\not=1$, $[kid_i]_h=[kid_1]_h$.
As shown in  Fig.~\ref{fig_flow_eq}, based on whether the kid widgets are line-wrapped, the Flow layout can be classified into two types: the wrap and the non-wrap versions.

{\bf Wrap Version: } 
The basic constraints for the wrap version are as follows:
\begin{itemize}
    \item the first kid widget $kid_1$ is located in the upper left corner of the flow container $w$: $[kid_1]_x=[w]_x$, $[kid_1]_y = [w]_y$.
    \item the last kid widget $kid_n$ is located within the flow container: $[kid_n]_x+[kid_n]_w\le[w]_x+[w]_w\wedge[kid_n]_y+[kid_n]_h\le [w]_y+[w]_h$.
    \item If the current row cannot hold the kid widget $kid_i$, $kid_i$ should be wrapped to the subsequent row. Otherwise, it should be located in the same row behind the previous widget $kid_{i-1}$. 
\end{itemize}

Specifically, the wrapping constraints are formalized as follows:
the condition that the current row cannot hold the kid widget $kid_i$ can be formalized as  $[kid_i]_w+[kid_{i-1}]_x+[kid_{i-1}]_w>[w]_x+[w]_w$, denoted as $e$. 
Wrapping the widget $kid_i$ to the next row can be formalized as $e\rightarrow ([kid_i]_x = [w]_x\wedge [kid_i]_y = [kid_{i-1}]_y+[kid_1]_h)$, indicating that $kid_i$ is located at the beginning of the next row.
Otherwise, the widget $kid_i$ is located behind $kid_{i-1}$ in the same row, which is formalized as $\neg e\rightarrow ([kid_i]_x = [kid_{i-1}]_x+[kid_{i-1}]_w\wedge [kid_i]_y = [kid_{i-1}]_y)$.

{\bf Non-wrap version: }
In contrast to the wrap version, if the kid widget cannot be contained within the current row, it will be invisible in the non-wrap version.
For example, as shown in Fig.~\ref{non-wrap}, $kid_3$ is invisible since it cannot be contained within the container.
Note that the non-wrap version of the Flow layout container is the only exception where not all kid widgets should be visible when the container is visible.
The basic constraints are as follows:
\begin{itemize}
    \item the first kid widget $kid_1$ is located in the upper left corner of the flow container $w$: $[kid_1]_x=[w]_x$, $[kid_1]_y = [w]_y$.
    \item each kid widget is compactly adjacent to the previous one in the same line: for $i\in[1,n)$ $[kid_{i+1}]_x=[kid_i]_x+[kid_i]_w$, $[kid_{i+1}]_y=[kid_i]_y$.
    \item those kid widgets located inside the container $w$ should be visible, while others should be invisible: for $kid_i\in [w]_k$, $[kid_i]_x+[kid_i]_w\le[w]_x+[w]_w\rightarrow [kid_i]_v$, $[kid_i]_x+[kid_i]_w>[w]_x+[w]_w\rightarrow \neg[kid_i]_v$
\end{itemize}

\subsubsection{Varying height widgets}
\begin{figure}[h]
  \centering
  \includegraphics[width=0.4\linewidth]{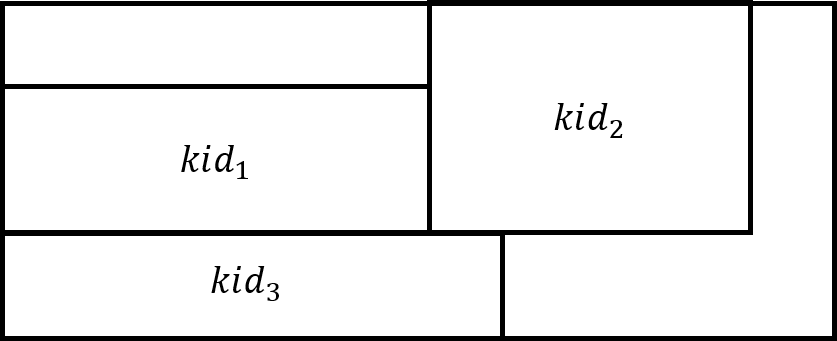}
  \caption{The wrap version of the flow container with varying height widgets: $kid_1,kid_2,kid_3$ have different heights, and $kid_1, kid_2$ are aligned at the bottom of the same row.}
  \label{flow_vary}
\end{figure}
Then we consider the more complicated scenario without the assumption that all kid widgets share the same height, indicating that the kid widgets have varying heights. Thus the position of widgets cannot be determined based on the preset height when wrapping lines, as shown in Fig.~\ref{flow_vary}.

It can also be classified into the wrap and non-wrap versions based on whether the kid widgets are line-wrapped.

In the {\bf wrap version}, The basic constraints are as follows:
\begin{itemize}
    \item the first widget is located on the left edge of the container: $[kid_1]_x=[w]_x$
    \item the last kid widget $kid_n$ is located within the flow container: $[kid_n]_y+[kid_n]_h\le [w]_y+[w]_h$, $[kid_n]_x+[kid_n]_w\le[w]_x+[w]_w$.
    \item If the current row cannot hold the kid widget $kid_i$, it should be wrapped to the next row. Otherwise, $kid_i$ is compactly adjacent to the previous one in the current row. Widgets in the same row should be aligned at the bottom.
\end{itemize}

To determine the position of widgets when wrapping, several auxiliary variables are introduced, namely $top_i$ and $max\_height_i$, representing the top position and the maximum height of the current row where the $i$th kid widget is located.
In the beginning, the auxiliary variable corresponding to the first kid widget $kid_1$ is initialized as $top_1=[kid_1]_y$ and $max\_height_1=[kid_1]_h$.
Then, we go through the kid widgets $kid_i$ where $i\in[2,n]$.
The condition that the current row cannot hold $kid_i$ is formalized as $[kid_i]_w+[kid_{i-1}]_x+[kid_{i-1}]_w>[w]_x+[w]_w$, denoted as $e$.

If $kid_i$ does not need to be wrapped to the next row, then it is compactly adjacent to the previous kid widget in the current row, and it is aligned with the previous one at the bottom: $\neg e\rightarrow([kid_i]_x=[kid_{i-1}]_x+[kid_{i-1}]_w\wedge[kid_i]_y+[kid_i]_h=[kid_{i-1}]_y+[kid_{i-1}]_h)$.
The auxiliary variables should also be updated accordingly:
the top position of the current row remains unchanged: $\neg e\rightarrow(top_i=top_{i-1})$;
the maximum height of the current row should be updated if the height of $kid_i$ is larger: 
$(\neg e\wedge [kid_i]_h>max\_heigt_{i-1})\rightarrow max\_heigt_i=[kid_i]_h$;
otherwise, it remains unchanged: $(\neg e\wedge [kid_i]_h\le max\_heigt_{i-1})\rightarrow max\_heigt_i=max\_heigt_{i-1}$.

If $kid_i$ needs to be wrapped to the next row, then the kid widget $kid_i$ should be kept on the left edge of the container, and located under the previous widget: $e\rightarrow([kid_i]_x=[w]_x\wedge[kid_i]_y=[kid_{i-1}]_y+[kid_{i-1}]_h)$.
The auxiliary variables should also be updated $e\rightarrow(top_i=[kid_i]_y\wedge max\_height_i=[kid_i]_h)$.
Moreover, the bottom position of the previous row is determined: $e\rightarrow ([kid_{i-1}]_y+[kid_{i-1}]_h=top_{i-1}+max\_height_{i-1})$.

The constraints of the {\bf Non-wrap} version are the same as the counterpart of the flow container with equal heights.

\subsection{Waterfall Layout}
\begin{figure}[h]
  \centering
  \includegraphics[width=0.25\linewidth]{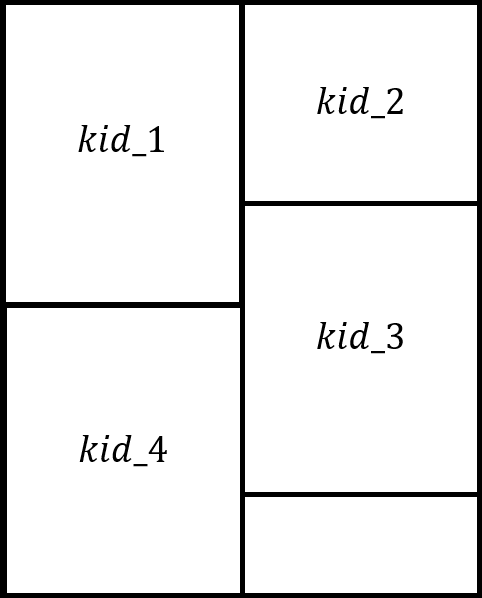}
  \caption{The waterfall layout container with 2 columns.}
  \label{waterfall}
\end{figure}

The waterfall layout container is widely applied in mobile App layout design.
This container consists of multiple columns, with each widget positioned at the top of the lowest column, as shown in Fig.~\ref{waterfall}.

{\bf Basic Constraints: }

Given a waterfall layout container $w$ with $c$ columns and $n$ kid widgets $kid_i\in [w]_k$, the constraints are as follows.
\begin{itemize}
    \item all kid widgets have the same width, which is $\frac{1}{c}$ of the width of the container: for $kid_i \in[w]_k,[kid_i]_w*c=[w]_w$.
    \item The kid widgets in the first row are aligned with the container at the top, and they are compactly adjacent to the previous one horizontally: for $i\in[1,c],[kid_i]_y=[w]_y$, $[kid_i]_x =(i-1)*[kid_1]_w+[w]_x $.
    \item Widgets not in the first row should be located in the column with the lowest current height.
\end{itemize}

To place the widgets not in the first row, several auxiliary variables named $hight\_i\_y$ are introduced, representing the current height of the $y$th column after placing the $i$th widget.

Based on the widgets in the first row, we first initialize the auxiliary variables after placing the $c$th widget: for $y\in [1,c]$, $hight\_{c}\_y = [kid_y]_h$.
Then, we go through the kid widget $kid_i$ where $i\in[c+1,n]$, and locate it in the $y$th column with the lowest current height.
Specifically, the condition that the $y$th column is the lowest when placing the $i$ can be formalized as $\bigwedge_{y'\not =y} (hight\_\{i-1\}\_y<hight\_\{i-1\}\_y')$, denoted as $l_y$.
When $l_y$ holds, $kid_i$ should be located in the $y$th column, which is formalized as $l_y\rightarrow([kid_i]_x = [w]_x+(y-1)*[kid_1]_w\wedge [kid_i]_y=low\_\{i-1\}\_y)$.
Moreover, the auxiliary variables should also be updated: $l_y\rightarrow(\bigwedge_{y'\not=y}hight\_{i}\_{y'}=low\_\{i-1\}\_{y'}\wedge hight\_{i}\_y=hight\_\{i-1\}\_y+[kid_i]_h)$, indicating that after placing the $i$th widget, the current height of the $y$th column needs to be updated, while others remain unchanged.

{\bf Available API: }
$set\_col(n)$ allows designers to define the column number as $n$.

\subsection{Table Layout}

\begin{figure}[h]
  \centering
  \includegraphics[width=0.35\linewidth]{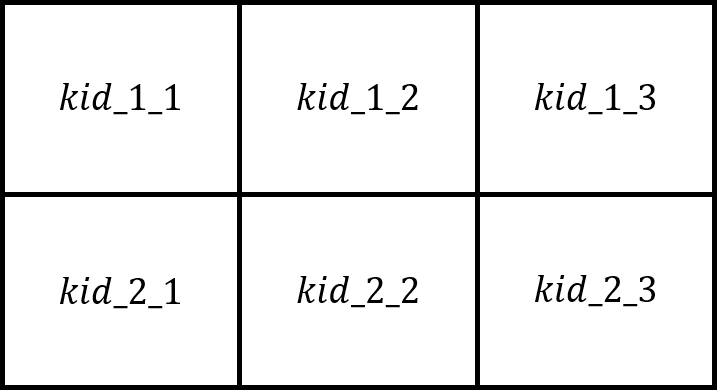}
  \caption{The Table layout container.}
  \label{table}
\end{figure}
Table layouts are frequently adopted in practical layout design, and the schematic diagram is shown in Fig.\ref{table}.
Given a table layout $w$ with $r$ rows and $c$ columns, we denote its kid widgets as $kid\_i\_j$ ($i\in[1,r],j\in[1,c]$), indicating that the widget is located in the $i$th row and $j$th column.

The basic constraints are as follows:
\begin{itemize}
    \item The first and last widgets are located in the left upper corner and right bottom corner respectively: $[kid\_1\_1]_x=[w]_x\wedge[kid\_1\_1]_y=[w]_y$, $[kid\_r\_c]_x+[kid\_r\_c]_w=[w]_x+[w]_w\wedge[kid\_r\_c]_y+[kid\_r\_c]_h=[w]_y+[w]_h$.
    \item widgets in the same row are aligned horizontally, and widgets in the same column are aligned vertically: given the $i$th row, for $j\not=1$, $[kid\_i\_1]_y=[kid\_i\_j]_y\wedge [kid\_i\_1]_h=[kid\_i\_j]_h$; analogously, given the $j$th column, for $i\not=1$, $[kid\_1\_j]_x=[kid\_i\_j]_x\wedge [kid\_1\_j]_w=[kid\_i\_j]_w$.
    \item In the first row and first column, each widget is compactly adjacent to the previous widget: for $j\in[1,c-1]$, $[kid\_1\_j]_x+[kid\_1\_j]_w=[kid\_1\_\{j+1\}]_x$; for $i\in[1,r-1]$, $[kid\_i\_1]_y+[kid\_i\_1]_h=[kid\_\{i+1\}\_1]_y$.
\end{itemize}

{\bf Available APIs: }

$set\_row(r)$ and $set\_col(c)$ specify the number of rows and columns respectively.

$set\_width(j, width)$ and $set\_height(i, height)$ allow designers to specify the width of the $j$th column and the height of the $i$th row.
These APIs will respectively add the following constraints: $[kid\_1\_j]_w=width$ and $[kid\_i\_1]_h=height$, and other columns(rows) remain to have the same width(height).
Without specification, all columns(rows) have the same width(height) on default: for $j\in[2,c]$, $[kid\_1\_j]_w=[kid\_1\_1]_w$, and for $i\in[2,r]$, $[kid\_i\_1]_h=[kid\_1\_1]_h$.

\subsection{Card Layout}
\begin{figure}[h]
  \centering
  \includegraphics[width=0.25\linewidth]{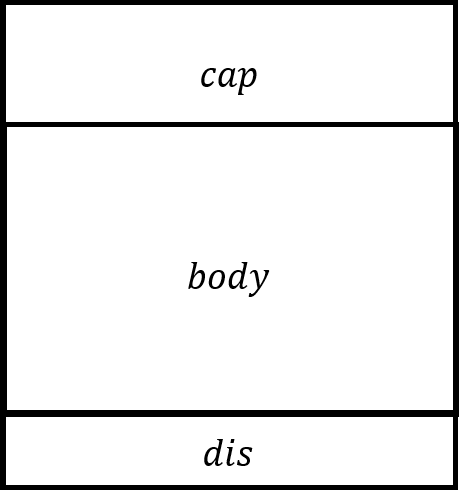}
  \caption{The Card layout container.}
  \label{card}
\end{figure}
The card layout container can be viewed as a special column layout container, as shown in Fig.~\ref{card}.
The card container $w$ has 3 kid widgets: the caption widget $cap$, the main body widget $body$, and the description widget $dis$.

The basic constraints are as follows:
\begin{itemize}
    \item All widgets are horizontally stretched to the container: for $kid_i\in [w]_k$, $[kid_i]_w=[w]_w\wedge [kid_i]_x=[w]_x $.
    \item The title and description are respectively kept to the upper and lower edges of the container: $[cap]_y=[w]_y$, $[dis]_y+[dis]_h=[w]_y+[w]_h$.
\end{itemize}

{\bf Available API:} 
$set\_proportion(i,p)$: designer can set the proportion of the height of $i$th kid widget to the height of the container: $[kid_i]_h = p*[w]_h$.

\subsection{Flex Layout}
We only consider the flex layout without wrapping lines, which can be regarded as an extended version of the Row \& Column layout container.
Without loss of generality, we consider the version with the horizontal main axis, and the vertical version can be analogized.
Several APIs are applied according to the CSS definition~\footnote{https://developer.mozilla.org/en-US/docs/Web/CSS/flex}:

{\bf Available APIs: }

The APIs concerning the alignment on the main axis and cross axis are as follows:
\begin{itemize}
    \item $strech()$: all kid widgets are stretched vertically on the cross axis: for $kid_i\in[w]_k$, $[kid_i]_y=[w]_y\wedge[kid_i]_h=[w]_h$.
    \item $space\_around()$: on the main axis, the margin between widgets is twice the edge padding: $[kid_1]_x-[w]_x = ([w]_x+[w]_w-[kid_n]_x-[kid_n]_w)$; for $i\in[2,n]$, $2*([kid_1]_x-[w]_x)=([kid_i]_x-[kid_{i-1}]_x-[kid_{i-1}]_w$.
    \item $space\_between()$: on the main axis, the first and last widget is kept on the border: $[kid_1]_x=[w]_x$, $[kid_n]_x+[kid_n]_w=[w]_x+[w]_w$; the margins between widgets are the same: for $i\in[3,n]$, $[kid_i]_x-[kid_{i-1}]_x-[kid_{i-1}]_w = [kid_2]_x-[kid_{1}]_x-[kid_{1}]_w$
    \item $flex\_start()$ \& $flex\_end()$: all widgets are aligned vertically to the start or end of the cross axis: for $kid_i\in[w]_k$, $[kid_i]_y=[w]_y$ or $[kid_i]_y+[kid_i]_h=[w]_y+[w]_h$.
    
\end{itemize}

$grow()$ \& $shrink()$: On the main axis, the designer can set the basic width of the $i$th widget, namely $basis_i$, and the proportion to grow or shrink, namely $flex\_grow_i$ and $flex\_shrink_i$.
If there is remaining space in the main axis: $[w]_w>\sum_{i\in[1,n]}basis_i$, denoted as $r$, then the remaining space is distributed proportionally among the widgets according to its $flex\_grow_i$: for $kid_i\in[w]_k$, $r\rightarrow ([kid_i]_w-basis_i)=\frac{flex\_grow_i}{\sum_{i\in[1,n]}flex\_grow_i}*([w]_w-\sum_{i\in[1,n]}basis_i)$.
Similarly, if there is no adequate space in the main axis: $[w]_w<\sum_{i\in[1,n]}basis_i$, denoted as $l$, then the widgets are shrunk proportionally according to their $flex\_shrink_i$: for $kid_i\in [w]_k$ $l\rightarrow (basis_i-[kid_i]_w)= \frac{flex\_shrink_i}{\sum_{i\in[1,n]}flex\_shrink_i}*(\sum_{i\in[1,n]}basis_i-[w]_w)$.

\section{Algorithm of Preview Module}
\label{appendix_a}

To preview the layout and debug these potential conflict constraints, the SMT solver is applied to solve these SMT formulas by traversing the range of the size properties, which can be the size properties of independent widgets or the screen width.
The Alg. \ref{preview} describes the previewing process in detail.

Given a preprocessed SMT formula and its original MaxSMT formula, we traverse the range of size property $p$ from the maximum value $max\_val$ to the minimum value $min\_val$ (Lines 2--12).
We first apply SMT solvers to solve the SMT formula by specifying the size property $p$ to the current value $curr\_p$ (Line 3).
Suppose the SMT formula is unsatisfiable at the current size property value (Line 4), the SMT solver is employed to solve the hard constraints of the original MaxSMT formula at the current size property value (Line 5).
If the hard constraints remain unsatisfiable (Line 6), indicating that the original specification contains conflict constraints at the current size, the ``unsatisfiable core'' function is performed to identify a conflicting constraint set and notify the designer to modify the specification (Line 7).
Otherwise, the current truth assignment of soft constraints is infeasible at the current size, indicating that the corresponding interval is discontinuous.
We extract the truth assignment of soft constraints related to the current value of the size property, denoted as $\alpha_{soft}$ (Line 9).
The {\it Interval-based soft constraints hardening} procedure should be reapplied to determine the continuous interval corresponding to $\alpha_{soft}$ (Line 10).
The MaxSMT formula is converted to a new SMT formula $F$ accordingly (Line 11).
After the traversing, the new SMT formula without conflict is returned (Line 13).
Since the overall SMT formula in the outer layer is based on the constraints of independent widgets, we first preview the SMT formulas in the inner layer.

\begin{algorithm}[t]
\caption{ Previewing process}
\label{preview}
\SetKwInOut{Input}{INPUT}
\SetKwInOut{Output}{OUTPUT}
\Input{The preprocessed SMT formulas $F$, the original MaxSMT formula $F_{max}$, and the size property $p$}
\Output{The conflict constraint sets or an SMT formula without conflict}
$curr\_p:=max\_val$\;
\While{$curr\_p\ge min\_val$}{
$SMT\_result:=$ apply SMT solver to solve $F\wedge(p==curr\_p)$\;
\If{$SMT\_result==unsat$}{
{\tcc{$F$ contains conflict constraints at the current size}}
$hard\_result:=$ apply SMT solver to solve $F_{max}.hard\wedge (p==curr\_p)$\;
\If{$hard\_result==unsat$}{
{\tcc{the original specification contains conflict hard constraints}}
{\bf return} the ``unsatisfiable core'' found by SMT solver\; 
}
\Else{
{\tcc{the interval corresponding to the current truth assignment of soft constraints is discontinuous}}
$\alpha_{soft}:=$ the assignment of soft constraints\;
$C\_new:=$ soft\_constraints\_hardening($F_{max}\wedge((\bigwedge_{\ell\in \alpha_{soft}}\ell)\rightarrow (p>curr\_p)),p$)\;
$F:=F_{max}.hard\wedge C\_new$\;
}
}
$curr\_p:= curr\_p-1$\;
}
{\bf return} $F$\;
\end{algorithm}

\end{appendices}
\bibliography{sn-bibliography}
\end{document}